\documentclass[twocolumn,journal, bookmarks=false]{IEEEtran}
\IEEEoverridecommandlockouts
\usepackage{color}
\usepackage{graphicx}
\usepackage{amsmath}
\usepackage{amssymb}
\usepackage{algorithm}
\usepackage{algorithmic}
\usepackage{amsmath}
\usepackage{multirow}
\usepackage{booktabs}
\usepackage{array}
\usepackage{amsthm}
\usepackage{stfloats}
\usepackage{caption}
\usepackage{subfigure}
\usepackage{bm}
\usepackage{setspace}
\usepackage{gensymb}
\usepackage{url}

\newtheorem{lemma}{Lemma}

\newcommand{\be}{\begin{equation}}
\newcommand{\ee}{\end{equation}}
\newcommand{\bea}{\begin{eqnarray}}
\newcommand{\eea}{\end{eqnarray}}
\newcommand{\ba}{\begin{array}}
\newcommand{\ea}{\end{array}}

\newcommand{\non}{\nonumber}

\renewcommand{\thesubsubsection}{\thesubsection.\arabic{subsubsection}}

\allowdisplaybreaks[4]

\title{Joint Waveform and Filter Designs for STAP-SLP-based MIMO-DFRC Systems
\thanks{R. Liu and M. Li are with the School of Information and Communication Engineering, Dalian University of Technology, Dalian 116024, China (e-mail: liurang@mail.dlut.edu.cn; mli@dlut.edu.cn).}
\thanks{Q. Liu is with the School of Computer Science and Technology, Dalian University of Technology, Dalian 116024, China (e-mail: qianliu@dlut.edu.cn).}
\thanks{A. L. Swindlehurst is with the center for Pervasive Communications and Computing, University of California, Irvine, CA 92697, USA (e-mail: swindle@uci.edu).}
}

\author{Rang Liu,~\IEEEmembership{Graduate Student Member,~IEEE,}
        Ming Li,~\IEEEmembership{Senior Member,~IEEE,}
        Qian Liu,~\IEEEmembership{Member,~IEEE,}\\
        and A. Lee Swindlehurst,~\IEEEmembership{Fellow,~IEEE}}

\begin{document}

\maketitle

\pagestyle{empty}
\thispagestyle{empty}

\begin{abstract}
Dual-function radar-communication (DFRC), which can simultaneously perform both radar and communication functionalities using the same hardware platform, spectral resource and transmit waveform, is a promising technique for realizing integrated sensing and communication (ISAC). Space-time adaptive processing (STAP) in multi-antenna radar systems is the primary tool for detecting moving targets in the presence of strong clutter. The idea of joint spatial-temporal optimization in STAP-based radar systems is consistent with the concept of symbol-level precoding (SLP) for multi-input multi-output (MIMO) communications, which optimizes the transmit waveform for each of the transmitted symbols. In this paper, we combine STAP and SLP and propose a novel STAP-SLP-based DFRC system that enjoys the advantages of both techniques. The radar output signal-to-interference-plus-noise ratio (SINR) is maximized by jointly optimizing the transmit waveform and receive filter, while satisfying the communication quality-of-service (QoS) constraint and various waveform constraints including constant-modulus, similarity and peak-to-average power ratio (PAPR). An efficient algorithm framework based on majorization-minimization (MM) and nonlinear equality constrained alternative direction method of multipliers (neADMM) methods is proposed to solve these complicated non-convex optimization problems. Simulation results verify the effectiveness of the proposed STAP-SLP-based MIMO-DRFC scheme and the associate algorithms.
\end{abstract}

\section{Introduction}\label{sec:introduction}

With the exponential growth of wireless services and the plethora of wireless devices, spectral resources are becoming increasingly scarce, motivating the urgent need for advanced spectrum sharing technologies.
The radar frequency bands, which have large portions of available spectrum, are promising candidates for sharing with various communication systems.
Spectrum sharing between radar and communication systems is consistent with the on-going convergence of integrated sensing and communication (ISAC) functions \cite{Zheng SPM 2019}-\cite{FLiu 2021}, and has led to substantial research interest in the coexistence, cooperation, and co-design of these two systems \cite{Liu TSP 18}-\cite{Zhang arxiv 2021}.

Unlike radar and communication coexistence/cooperation (RCC) which requires interference management and side-information exchange, ISAC enables the co-design of the sensing and communication functions.
By efficiently sharing the same spectral resources, a given hardware platform, and a joint signal processing framework, ISAC can realize both sensing and communication functions and even result in mutual benefits \cite{Cui 2021}-\cite{An Liu 2021}.
ISAC can provide considerable gains in terms of spectral/energy/hardware/cost efficiency and has attracted significant research interest in both academia and industry. It is believed that ISAC will become a promising technology in future wireless communication systems to support various application scenarios, such as vehicular networks, smart cities, environmental monitoring, remote sensing, internet-of-things (IoT), etc. Moreover, ISAC is also considered to be a key enabling technology for next-generation cellular and Wi-Fi systems.

ISAC systems can have different levels of integration.
In a looser configuration, these two functions are just physically integrated on a given platform, but employ different sets of hardware components and/or transmit waveforms. This loose integration may only offer limited benefits such as lower signalling overhead and better interference management.
The rationale of ISAC is that sensing and communication functions are tightly integrated and can be simultaneously performed using a fully-shared transmitter/receiver, the same frequency bands, and the same dual-functional waveforms, which allows for significantly greater improvements in efficiency \cite{Cui 2021}.
An ISAC system with such tight integration is more often referred to as a dual-functional radar-communication (DFRC) system in the literature \cite{Liu TWC 18}-\cite{Cheng TCCN 2021}. The main goal of a DFRC system is to generate novel dual-functional waveforms to simultaneously perform radar sensing and communication functions.
Multi-input multi-output (MIMO) architectures have been widely employed in DFRC systems to improve the spatial-domain waveform diversity for radar sensing \cite{Li SPM 2007}, as well as to achieve beamforming gains and spatial multiplexing for multi-user communications.
Thanks to advancements in fully co-designed radar sensing and communication waveforms, MIMO-DFRC is recognized as a key enabler for ISAC systems to significantly improve spectral efficiency, reduce device size, cost and power consumption, and enhance performance, all of which are the focus of this paper.

Due to the inherently conflicting requirements of radar sensing and communication functions, transmit waveform design is pivotal in pursuing better performance trade-offs for MIMO-DFRC systems \cite{Zhang 2021}.
Therefore, many researchers have devoted their study to transmit waveform designs with various radar sensing and communication metrics \cite{Liu TWC 18}-\cite{Cheng TCCN 2021}, \cite{Qian TSP 2018}-\cite{Yuan TWC 2020}, e.g., average transmit beampattern, signal-to-interference-plus-noise ratio (SINR), mutual information, similarity with a reference waveform, achievable rate, etc.
However, most existing research for implementing MIMO-DFRC only focuses on designing the spatial second-order statistics of the transmit waveforms and ignores the Doppler bin in the temporal domain.
Moreover, an overly simplified radar sensing environment without clutter is usually assumed in the existing literature.
Therefore, the target detection performance of these designs may be not satisfactory, and could even be unacceptable in a hostile radar sensing environment.

Space-time adaptive processing (STAP) is an effective technique for achieving adaptive clutter suppression and better target detection in multi-antenna radars \cite{Guerci 2014}-\cite{Wu TSP 2018}. STAP optimizes the spatial-temporal transmit waveforms rather than their spatial second-order statistics to maximize the output SINR performance, for the purpose of suppressing the clutter. Since the waveform optimization exploits degrees of freedom (DoFs) in both the spatial and temporal domains, the performance of identifying a target in the presence of strong clutter over widely spread ranges and angular regions is significantly improved. Using a priori knowledge about the clutter, e.g., the clutter covariance matrix (CCM) \cite{Sun GRSL  2018}-\cite{Sun GRSL}, there has been growing interest in designing the waveforms using knowledge-aided techniques to improve the STAP performance.

In communications applications, the recently emerged symbol-level precoding (SLP) technique also exploits available DoFs in both the spatial and temporal domains to improve link performance.
In particular, SLP designs the transmit precoder in each time slot (i.e., the transmit waveform samples) based on the specific transmitted symbols themselves rather than their second-order statistics. The transmit precoder/waveform can be designed to convert harmful interference into beneficial signal power, and such constructive interference (CI) can improve the communication quality-of-service (QoS)  \cite{Masouros TWC 2009}-\cite{Liu TWC 2021}. The flexibility offered by SLP in the time domain and its ability to achieve better communication QoS make SLP techniques a promising candidate for MIMO-DFRC systems, in which the transmit waveform used for radar sensing simultaneously carries information symbols for wireless communications.

Very limited research has been conducted to exploit SLP for joint radar-communication systems.
In prior work \cite{Liu TSP 18}, the SLP technique was employed in an RCC system to take advantage of interference exploitation.
However, in RCC systems, the SLP design only optimizes the communication performance metric while simultaneously suppressing interference to the radar system regardless of the specific radar waveforms.
Another very recent work \cite{Liu JSTSP 2021} introduces SLP to  MIMO-DFRC systems for the first time and illustrates that SLP can provide more accurate angle estimation and better target detection performance, as well as lower symbol-error-rate (SER) for multi-user communications compared with conventional block-level precoding (BLP) schemes.
Nevertheless, this work only optimizes the transmit waveform based on the beampattern similarity metric, and does not consider the properties of the radar waveform in the temporal-domain.
The spatial-temporal receive filter and clutter suppression are also not taken into account. Therefore, the flexibility of SLP has not been fully exploited for MIMO-DFRC systems in the prior literature.

Motivated by the above discussion, in this paper we leverage STAP and CI-based SLP techniques for implementing MIMO-DFRC to combine their advantages for both radar and communication functions.
In particular, we consider a multi-antenna base station (BS) that simultaneously uses active sensing to detect a target in the presence of strong signal-dependent clutter and transfers information symbols to multiple single-antenna users.
The transmit waveform and receive filter of the BS are jointly optimized to maximize the radar output SINR under communication QoS constraints and several different radar waveform constraints.
The main contributions can be summarized as follows:

\begin{itemize}
\item For the first time, we integrate STAP and SLP techniques to implement MIMO-DFRC in order to achieve considerable improvements in target detection performance in the presence of strong clutter, as well as to boost multi-user communication performance by converting harmful interference into beneficial signal power. Compared with conventional BLP-based waveform designs, STAP and SLP techniques impose quite different radar sensing and communication constraints in the temporal domain and consequently result in brand-new waveform design problems.

\item We first model the joint transmit waveform and receive filter optimization problem for the novel STAP-SLP-based MIMO-DFRC system.
    Then, focusing on the constant-modulus constrained waveform design problem, we employ the majorization-minimization (MM) method and derive a tractable surrogate function, and then exploit the novel nonlinear equality constrained alternative direction method of multipliers (neADMM) method to convert the problem into manageable sub-problems. Detailed derivations and efficient algorithms are developed to obtain the optimal solutions for each sub-problem.

\item Next, we generalize the proposed MM-neADMM algorithm to waveform designs with constant-modulus similarity and peak-to-average power ratio (PAPR) constraints.

\item We provide extensive simulation results to verify the advantages of jointly exploiting STAP and CI-based SLP techniques to implement MIMO-DFRC and demonstrate the effectiveness of the proposed algorithms under different waveform constraints.
\end{itemize}


\textit{Notation}:
Boldface lower-case and upper-case letters indicate column vectors and matrices, respectively.
$(\cdot)^T$ and $(\cdot)^H$ denote the transpose and the transpose-conjugate operations, respectively.
$\mathbb{C}$ and $\mathbb{R}$ denote the sets of complex numbers and real numbers, respectively.
$| a |$, $\|\mathbf{a}\|$, and $\|\mathbf{a}\|_\infty$ are the magnitude of a scalar $a$, the 2-norm of a vector $\mathbf{a}$, and the infinity norm of a vector $\mathbf{a}$, respectively.
$\angle{a}$ is the angle of complex-valued $a$.
$\mathfrak{R}\{a\}$ and $\mathfrak{I}\{a\}$ denote the real and imaginary part of a scalar $a$, respectively.
$\otimes$ denotes the Kronecker product.
$\mathbb{E}\{\cdot\}$ denotes the expectation operation.
$\text{Tr}\{\mathbf{A}\}$ takes the trace of a matrix $\mathbf{A}$ and $\text{vec}\{\mathbf{A}\}$ vectorizes the matrix $\mathbf{A}$.
$\mathbf{I}_M$ indicates an $M\times M$ identity matrix.

\section{System Model and Problem Formulation}\label{sec:system model}

\begin{figure}[!t]
\centering
\includegraphics[width = 2.7 in]{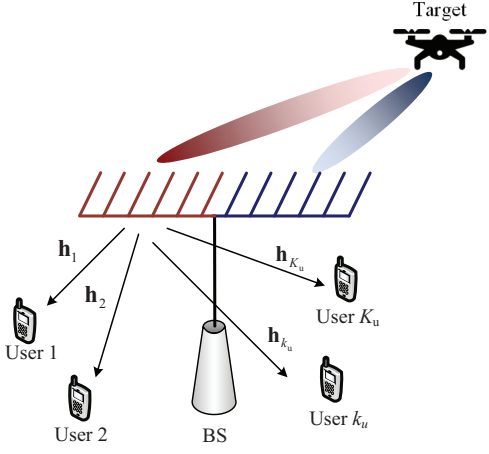}\vspace{0.1 cm}
\caption{The considered MIMO-DFRC system.}
\label{fig:system_model}\vspace{0.2 cm}
\end{figure}

Consider a colocated narrowband DFRC system as shown in Fig. \ref{fig:system_model}, where a BS is equipped with $N_\text{t}$ transmit antennas and $N_\text{r}$ receive antennas arranged as uniform linear arrays (ULAs) with antenna spacing $d_\text{t}$ and $d_\text{r}$, respectively.
The BS aims to detect a target in the presence of strong signal-dependent clutter and simultaneously provide downlink wireless communication services to $K_\text{u}$ single-antenna users.
\textcolor{black}{In order to detect moving targets in the presence of strong signal-dependent clutter over widely spread ranges, angular regions and Doppler frequencies, the BS uses STAP to exploit all available DoFs in both the spatial and temporal domains.
In other words, the BS, which can be thought of as a colocated MIMO radar, uses STAP to jointly design the spatial-temporal transmit waveform and receive filter for achieving better target detection and clutter suppression performance.}
Meanwhile, in order to simultaneously realize satisfactory communication performance, the information symbols are carried by the same transmit waveforms using CI-based SLP for better communication QoS.

We assume that the radar is interrogating a moving target at the azimuth direction $\theta_0$ with a speed $v_0$, in the presence of strong clutter from neighboring range cells.
Assume that a burst of $M$ pulses is transmitted from the radar transmitter in a coherent processing interval (CPI) with a constant pulse repetition frequency (PRF) $f_\text{r}$, and hence a constant pulse repetition interval (PRI) $T_\text{r} = 1/f_\text{r}$.
Let $\widetilde{x}_{n_\text{t}}(t)$ be the transmitted waveform of the $n_\text{t}$-th transmit antenna, $n_\text{t} = 1, \ldots, N_\text{t}$, and define $\widetilde{\mathbf{x}}(t) \triangleq [\widetilde{x}_1(t),\ldots,\widetilde{x}_{N_\text{t}}(t)]^T$.
It should be emphasized that, due to the use of SLP, the transmit waveforms vary from pulse-to-pulse to realize the multi-user communications, unlike what is assumed in most conventional MIMO radar systems that repeatedly transmit the same pulse.

\subsection{Radar Received Signal and Radar Performance Metric}

The radar received signal $\mathbf{y}$ in the cell under test (CUT) can be expressed as one of two possible hypotheses:
\begin{equation}\label{eq:binary hypothesis}
\left\{ \begin{array}{l l }
\mathcal{H}_0: \mathbf{y} = \mathbf{y}_\text{c} + \mathbf{z},\\
\mathcal{H}_1: \mathbf{y} = \mathbf{y}_0 + \mathbf{y}_\text{c} + \mathbf{z},\end{array} \right.
\end{equation}
where $\mathbf{y}_0$ and $\mathbf{y}_\text{c}$ respectively represent the received signal returns reflected from the target and the clutter, and the vector $\mathbf{z}\sim \mathcal{CN}(0,\sigma_\text{r}^2\mathbf{I})$ denotes additive white Gaussian noise (AWGN) at the receive antennas.
The radar system decides whether a target is observed by testing the binary hypotheses in~(\ref{eq:binary hypothesis}).

The received echo from the target with direction of arrival (DoA) $\theta_0$ can be written as
\be
\widetilde{\mathbf{y}}_0(t) = \alpha_0\mathbf{b}(\theta_0)\mathbf{a}^H(\theta_0)\widetilde{\mathbf{x}}(t-\tau_0)e^{j2\pi(f_0+f_\text{d})(t-\tau_0)},
\ee
where $\alpha_0$ represents the target amplitude with $\mathbb{E}\big\{|\alpha_0|^2\big\} = \sigma_0^2$, the scalar $\tau_0$ is the two-way propagation delay, $f_0$ is the carrier frequency of the transmit waveform, and $f_\text{d} = 2v_0/\lambda$ is the target Doppler frequency with $\lambda = c/f_0$ denoting the wavelength and  $c$ representing the speed of light.
The vectors $\mathbf{b}(\theta)$ and $\mathbf{a}(\theta)$ are the steering vectors for the receive and transmit signals at angle $\theta$, respectively:
\begin{subequations}\begin{align}
\mathbf{b}(\theta) &\triangleq \big[1,e^{\jmath2\pi f_\text{s}},\ldots,e^{-\jmath2\pi(N_\text{r}-1)f_\text{s}}\big]^T,\\
\mathbf{a}(\theta) &\triangleq \big[1,e^{\jmath2\pi f_\text{s}d_\text{t}/d_\text{r}},\ldots,e^{-\jmath2\pi(N_\text{t}-1)f_\text{s}d_\text{t}/d_\text{r}}\big]^T,
\end{align}\end{subequations}
where $f_\text{s}\triangleq d_\text{r}\sin\theta/\lambda$ denotes the normalized spatial frequency.

The received signal is first down-converted to baseband and then passed through an analog-to-digital converter.
For simplicity, we absorb the constant phase terms associated with $\tau_0$ into the target amplitude and assume the intra-pulse Doppler shift is negligible.
Thus, for the $m$-th pulse, the baseband digital samples at the considered range gate can be expressed in matrix form as
\be\label{eq:target baseband signal}
\mathbf{Y}_{0,m} = \alpha_0 e^{j2\pi(m-1)f_\text{d}T_\text{r}}\mathbf{b}(\theta_0)\mathbf{a}^H(\theta_0)\mathbf{X}_m,
\ee
where $\mathbf{X}_m = [\widetilde{\mathbf{x}}_{m,1},\ldots,\widetilde{\mathbf{x}}_{m,N_\text{t}}]^T\in\mathbb{C}^{N_\text{t}\times N}$ denotes the waveform matrix, and $\widetilde{\mathbf{x}}_{m,n_\text{t}}\in\mathbb{C}^N$ is a vector of samples of $\widetilde{x}_{n_\text{t}}(t)$, where $N$ samples are taken per pulse.
Then, we vectorize the received baseband digital samples in a CPI by letting $\mathbf{y}_0 = [\text{vec}\{\mathbf{Y}_{0,1}^T\}^T,\ldots,\text{vec}\{\mathbf{Y}_{0,M}^T\}^T]^T$, which can be expressed as
\be
\mathbf{y}_0 = \alpha_0\overline{\mathbf{X}}\big(\mathbf{d}(f_\text{d})\otimes\mathbf{b}(\theta_0)\otimes\mathbf{a}(\theta_0)\big),
\ee
where
\be\label{eq:Xbar}
\overline{\mathbf{X}} \triangleq \text{blkdiag}\big\{\mathbf{I}_{N_\text{r}}\otimes \mathbf{X}_1^T,\ldots,\mathbf{I}_{N_\text{r}}\otimes \mathbf{X}_M^T\big\},
\ee
and $\mathbf{d}(f_\text{d}) \triangleq [1,\ldots,e^{j2\pi(M-1)f_\text{d}T_\text{r}}]^T$ denotes the Doppler response vector.
\textcolor{black}{For simplicity, we define the spatial-temporal steering vector as
\be\label{eq:define u}
\mathbf{u}(f_\text{d},\theta) \triangleq \mathbf{d}(f_\text{d})\otimes\mathbf{b}(\theta)\otimes\mathbf{a}(\theta).
\ee}
The spatial-temporal steering vector of the target of interest is denoted as $\mathbf{u}_0 = \mathbf{u}(f_\text{d},\theta_0)$.
Thus, the received signal vector $\mathbf{y}_0$ can be re-written in a concise form as
\be
\mathbf{y}_0 = \alpha_0\overline{\mathbf{X}}\mathbf{u}_0.
\ee

In addition to the signal reflected from the target, the radar receiver simultaneously receives unwanted clutter reflected by trees, tall buildings, cars, etc., which are generally spread in both the spatial (e.g., azimuth and range) and Doppler dimensions.
Since the signal-dependent clutter is possibly stronger than the target return and deteriorates the target detection performance, it should be carefully taken into consideration in waveform designs.
We assume that the clutter is generated from the CUT and $2L$ other adjacent range cells, each of which is approximated by $N_\text{c}$ clutter patches randomly distributed in azimuth.
The origin of the range coordinates is set at the target range bin.

Similar to (\ref{eq:target baseband signal}), for the $m$-th pulse, the baseband digital samples of the return from the $k$-th clutter patch in the $l$-th range cell can be written as
\be
\mathbf{Y}_{\text{c},m,l,k} = \alpha_{\text{c},l,k} e^{j2\pi(m-1)f_{\text{c},l,k}T_\text{r}}\mathbf{b}(\theta_{\text{c},l,k})\mathbf{a}^H(\theta_{\text{c},l,k})\mathbf{X}_m\mathbf{J}_l,
\ee
where $\alpha_{\text{c},l,k}$, $f_{\text{c},l,k}$, and $\theta_{\text{c},l,k}$ respectively denote the amplitude, Doppler frequency, and DoA of the $k$-th clutter patch in the $l$-th range cell, $k = 1, \ldots, N_\text{c}$, $l = -L, \ldots, L$, and $\mathbb{E}\big\{|\alpha_{\text{c},l,k}|^2\big\} = \sigma_\text{c}^2$.
The shift matrix $\mathbf{J}_l\in\mathbb{R}^{N\times N}$ is defined by
\begin{equation}
\mathbf{J}_l(i,j) = \left\{
             \begin{array}{lr}1,~~~i-j+l=0,&\\
             0, ~~~\text{otherwise},&
             \end{array}
\right.
\end{equation}
and $\mathbf{J}_{-l} = \mathbf{J}_l^T$.
The received signal vector from the $k$-th clutter patch in the $l$-th range cell $\mathbf{y}_{\text{c},l,k} =  [\text{vec}\{\mathbf{Y}_{\text{c},1,l,k}^T\}^T,\ldots,\text{vec}\{\mathbf{Y}_{\text{c},M,l,k}^T\}^T]^T$ can be expressed as
\be
\mathbf{y}_{\text{c},l,k} = \alpha_{\text{c},l,k} \big(\mathbf{I}_{N_\text{r}}\otimes\mathbf{I}_M\otimes\mathbf{J}_l^T)\overline{\mathbf{X}}
\mathbf{u}(f_{\text{c},l,k},\theta_{\text{c},l,k}).
\ee
Defining $\overline{\mathbf{J}}_l\triangleq\mathbf{I}_{N_\text{r}}\otimes\mathbf{I}_M\otimes\mathbf{J}_l^T$ and \textcolor{black}{the spatial-temporal steering vector $\mathbf{u}_{\text{c},l,k}\triangleq \mathbf{u}(f_{\text{c},l,k},\theta_{\text{c},l,k})$ as given in (\ref{eq:define u})}, the clutter returns from all the clutter patches can be expressed as
\be
\mathbf{y}_\text{c} = \sum^L_{l=-L}\sum_{k=1}^{N_\text{c}}\mathbf{y}_{\text{c},l,k} = \sum^L_{l=-L}\sum_{k=1}^{N_\text{c}}\alpha_{\text{c},l,k}\overline{\mathbf{J}}_l\overline{\mathbf{X}}\mathbf{u}_{\text{c},l,k}.
\ee
The clutter covariance matrix (CCM) is thus given by
\be
\mathbf{R}_\text{c} = \mathbb{E}\{\mathbf{y}_\text{c}\mathbf{y}_\text{c}^H\} = \sum^L_{l=-L}\overline{\mathbf{J}}_l\overline{\mathbf{X}}\mathbf{M}_l\overline{\mathbf{X}}^H\overline{\mathbf{J}}_l^H,
\ee
where the inner CCM for the $l$-th range cell is defined by \cite{Sun TGRS 2021}
\be\label{eq:Ml}
\mathbf{M}_l = \mathbb{E}\Big\{\sum_{k=1}^{N_\text{c}}|\alpha_{\text{c},l,k}|^2
\mathbf{u}_{\text{c},l,k}\mathbf{u}_{\text{c},l,k}^H\Big\}.
\ee

We note that some prior work such as \cite{Tang TSP 2016}-\cite{Tang TSP 2020}, \cite{Wu TSP 2018} is developed based on the assumption that the spatial-temporal steering vectors $\mathbf{u}_{\text{c},l,k}$ of the clutter are known a priori, \textcolor{black}{or in other words that the azimuths, ranges, and Doppler frequencies of the clutter patches are exactly known. In practice, however, it is difficult to obtain parameters such as these for the signal-dependent clutter. Thus, in this paper, we assume that only the inner CCMs $\mathbf{M}_l$ must be known (or estimated from the data), which is a more realistic assumption \cite{Sun GRSL 2018}-\cite{Sun GRSL}.}

Denote $\mathbf{w}\in\mathbb{C}^{MNN_\text{r}}$ as the associated linear spatial-temporal receive filter whose output can be expressed as
\begin{subequations}\begin{align}
r &= \mathbf{w}^H \mathbf{y} = \mathbf{w}^H \left( \mathbf{y}_0 + \mathbf{y}_c + \mathbf{z} \right) \\
&= \alpha_0\mathbf{w}^H\overline{\mathbf{X}}\mathbf{u}_0 + \mathbf{w}^H\sum_{l=-L}^L\sum_{k=1}^{N_\text{c}}\alpha_{\text{c},l,k}\overline{\mathbf{J}}_l
\overline{\mathbf{X}}\mathbf{u}_{\text{c},l,k} + \mathbf{w}^H\mathbf{z}.
\end{align}\end{subequations}
Thus, the radar output SINR is given by
\begin{equation}\label{eq:gamma}
\text{SINR} = \frac{\sigma_0^2|\mathbf{w}^H\overline{\mathbf{X}}\mathbf{u}_0|^2}
{\mathbf{w}^H\big[\sum_{l=-L}^L\overline{\mathbf{J}}_l\overline{\mathbf{X}}\mathbf{M}_l
\overline{\mathbf{X}}^H\overline{\mathbf{J}}_l^H+\sigma_\text{r}^2\mathbf{I}\big]\mathbf{w}}.
\end{equation}
\textcolor{black}{In order to facilitate the discrimination between the two hypotheses in (\ref{eq:binary hypothesis}) for better target detection performance, the joint transmit waveform and receive filter design problem from the radar perspective aims to maximize the radar output SINR (\ref{eq:gamma}).}

In maximizing the SINR (\ref{eq:gamma}), the transmit radar waveforms are subject to certain constraints due to hardware limitations and  other radar sensing requirements.
For notational simplicity, we define the waveform matrix in a CPI as $\mathbf{X} \triangleq [\mathbf{X}_1, \ldots, \mathbf{X}_M]$, and the waveform vector $\mathbf{x} \triangleq \text{vec}\{\mathbf{X}\}$, $\mathbf{x} = [x_1,\ldots,x_{MNN_\text{t}}]^T$.
To achieve the best possible performance, we assume the total power constraint is satisfied with equality:
\be\label{eq:total power constraint}
\left\|\mathbf{x}\right\|^2 = P,
\ee
where $P$ is the total available transmit power for a CPI.
Considering the hardware requirements, constant-modulus waveforms are more preferred in practical radar systems to avoid nonlinear distortion:
\be\label{eq:CE constraint}
|x_i| = \sqrt{P/(MNN_\text{t})},~~\forall i = 1,\ldots,MNN_\text{t}.
\ee

The PAPR constraint compromises the strict constant-modulus constraint (\ref{eq:CE constraint}) by allowing power variation within a certain level and can provide a higher radar output SINR.
The PAPR and its constraint are usually defined by
\be\label{eq:PAR}
\text{PAPR} = \frac{\underset{1\leq i\leq MNN_\text{t}}\max\{|x_i|^2\}}{\|\mathbf{x}\|^2/(MNN_\text{t})}\leq 1+\varepsilon,
\ee
where $\varepsilon>0$ is a predefined parameter to control the PAPR level.
Substituting the total power constraint (\ref{eq:total power constraint}) into (\ref{eq:PAR}), the PAPR constraint can be re-written as
\be\label{eq:PAR constraint}
|x_i| \leq \sqrt{(1+\varepsilon) P/(MNN_\text{t})},~~\forall i = 1,\ldots,MNN_\text{t}.
\ee

Moreover, similarity between the designed waveform and a reference waveform may also be necessary to achieve some desired ambiguity function or pulse compression properties.
If we denote the reference waveform as $\mathbf{x}_0\triangleq [x_{0,1},\ldots,x_{0,MNN_\text{t}}]^T$, the similarity constraint is formulated as
\be\label{eq:similarity}
\left\|\mathbf{x}_0-\mathbf{x}\right\|_\infty \leq \xi,
\ee
where $\xi$ determines the level of allowable similarity deviation.
Alternatively, the similarity constraint can be re-written as
\be\label{eq:similarity constraint}
|x_{0,i}-x_i| \leq \xi,~~\forall i = 1,\ldots,MNN_\text{t}.
\ee

\subsection{User Received Signal and Communication Performance Metric}
In addition to its radar sensing function, the BS also attempts to simultaneously deliver information symbols to $K_\text{u}$ single-antenna users using the same transmit waveform.
In particular, we denote the symbol vector to be transmitted by the $n$-th sample of the $m$-th pulse as  $\mathbf{s}_{m,n} \triangleq [s_{m,n,1},\ldots,s_{m,n,K_\text{u}}]^T$, where each symbol is assumed to be independently selected from an $\Omega$-phase shift keying (PSK) constellation.
The waveform sample $\mathbf{x}_{m,n}$, which is the $n$-th column of $\mathbf{X}_m$, must be designed to carry the $K_\text{u}$ different information symbols in $\mathbf{s}_{m,n}$.
The corresponding received signal at the $k_\text{u}$-th user can be expressed as
\be
r_{m,n,k_\text{u}} = \mathbf{h}_{k_\text{u}}^H\mathbf{x}_{m,n} + z_{m,n,k_\text{u}},
\ee
where $\mathbf{h}_{k_\text{u}} \in \mathbb{C}^{N_\text{t}}$ represents the Rayleigh fading channel from the BS to the $k_\text{u}$-th user, and $z_{m,n,k_\text{u}}\sim\mathcal{CN}(0,\sigma^2)$ is AWGN at the $k_\text{u}$-th user.
The nonlinear mapping from $\mathbf{s}_{m,n}$ to $\mathbf{x}_{m,n}$ is achieved by the CI-based SLP design as briefly presented below.

\begin{figure}[!t]
\centering
\includegraphics[width = 2.4 in]{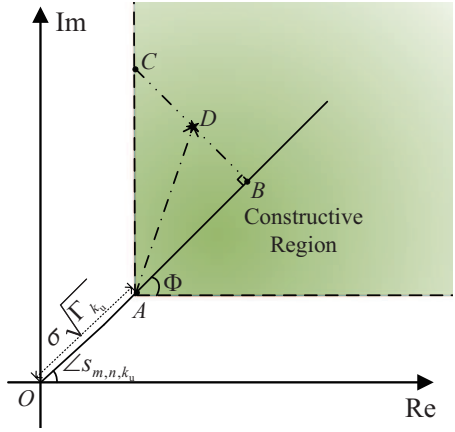}\vspace{0.2 cm}
\caption{CI-based SLP for QPSK constellation.}\label{fig:CR}
\end{figure}

We use the quadrature-PSK (QPSK) constellation (i.e., $\Omega = 4$) as an example to illustrate the CI-based SLP approach as shown in Fig.~\ref{fig:CR}, where $\Phi = \pi/\Omega$ is half of the angular range of the decision regions.  Fig.~\ref{fig:CR} shows the case where the desired symbol of the $k_\text{u}$-th user is  $(1/\sqrt{2},\jmath/\sqrt{2})$, whose decision boundaries are the positive halves of $x$ and $y$ axes. Assume point $D$ denotes the received noise-free signal $\widetilde{r}_{m,n,k_\text{u}} = \mathbf{h}_{k_\text{u}}^H\mathbf{x}_{m,n}$.
Unlike conventional block-level precoding approaches aiming to eliminate the interference, the CI-based SLP approach attempts to exploit known symbol information to convert the multi-user interference into constructive components, which can enhance the communication QoS.
In particular, let $\Gamma_{k_\text{u}}$ be the QoS requirement of the $k_\text{u}$-th user.
If the interference is entirely eliminated, the received noise-free signal should be at point $A$ to satisfy $\widetilde{r}_{m,n,k_\text{u}} = \sigma\sqrt{\Gamma_{k_\text{u}}}s_{m,n,k_\text{u}}$, i.e., $ \big|\widetilde{r}_{m,n,k_\text{u}}\big|^2/\sigma^2 = \Gamma_{k_\text{u}}$.
Instead of interference elimination/suppression, the CI-based SLP approach can utilize the interference to push the received noise-free signal deeper into the corresponding constructive (green) region, where the QoS requirement $\Gamma_{k_\text{u}}$ is guaranteed and the distance between the received noise-free signal and its decision boundaries is further enlarged. Thus, lower SER and better QoS can be achieved using this CI-based SLP approach.

The relationship governing the definition of the constructive region can be geometrically expressed as $|\overrightarrow{BC}| - |\overrightarrow{BD}| \geq 0$.
Due to space limitations, we omit the derivations and recommend the readers to \cite{MA ICST 2018}-\cite{Liu TWC 2021} for details. The QoS constraints that guarantee that the noise-free received signal $\widetilde{r}_{m,n,k_\text{u}}$ lies in the constructive region can be expressed as
\be\label{eq:communication QoS}\begin{aligned}
&\Re\big\{\mathbf{h}_{k_\text{u}}^H\mathbf{x}_{m,n}e^{-\jmath\angle{s_{m,n,k_\text{u}}}}-\sigma\sqrt{\Gamma_{k_\text{u}}}\big\}
   \sin \Phi \\
&~~~- \big|\Im\big\{\mathbf{h}_{k_\text{u}}^H\mathbf{x}_{m,n}e^{-\jmath\angle{s_{m,n,k_\text{u}}}}\big\}\big|\cos\Phi
   \geq 0,~\forall k_\text{u}, m, n.
\end{aligned}\ee
In order to represent (\ref{eq:communication QoS}) in a compact form, we define
\be\non\begin{aligned}
\widetilde{\mathbf{h}}_{(2{k_\text{u}}-2)MN+j}^H &\triangleq \mathbf{e}_{j,MN}^T\hspace{-0.08cm}\otimes\hspace{-0.08cm}\mathbf{h}_{k_\text{u}}^H e^{-\jmath\angle s_{m,n,k_\text{u}}}(\sin\Phi\hspace{-0.08cm}+\hspace{-0.08cm}
 e^{-\jmath\frac{\pi}{2}}\cos\Phi),\\
\widetilde{\mathbf{h}}_{(2{k_\text{u}}-1)MN+j}^H &\triangleq \mathbf{e}_{j,MN}^T\hspace{-0.08cm}\otimes\hspace{-0.08cm}\mathbf{h}_{k_\text{u}}^H e^{-\jmath\angle s_{m,n,k_\text{u}}}(\sin\Phi\hspace{-0.08cm}-\hspace{-0.08cm}
 e^{-\jmath\frac{\pi}{2}}\cos\Phi),\\
\gamma_{(2k_\text{u}-2)MN+j} &\triangleq \sigma\sqrt{\Gamma_{k_\text{u}}}\sin\Phi,\\
\gamma_{(2k_\text{u}-1)MN+j} &\triangleq \sigma\sqrt{\Gamma_{k_\text{u}}}\sin\Phi,
\end{aligned}\ee
where the vector $\mathbf{e}_{j,MN}\in\mathbb{R}^{MN}$ indicates the $j$-th column of an $MN\times MN$ identity matrix.
Then, the communication QoS constraints are equivalently re-written as
\be\label{eq:communication constraint}
\Re\big\{\widetilde{\mathbf{h}}_i^H\mathbf{x}\big\}\geq \gamma_i,~~\forall i = 1, \ldots, 2K_\text{u}MN.
\ee


\subsection{Problem Formulation}

In this paper, we aim to jointly design the transmit waveform $\mathbf{x}$ and the receive filter $\mathbf{w}$ to maximize the radar output SINR (\ref{eq:gamma}), while satisfying the communication QoS requirements (\ref{eq:communication constraint}), the total power constraint (\ref{eq:total power constraint}), and one of the waveform constraints (\ref{eq:CE constraint}), (\ref{eq:PAR constraint}), or (\ref{eq:similarity constraint}).
Therefore, the optimization problem is formulated as
\begin{subequations}\label{eq:original problem}
\begin{align}
&\underset{\mathbf{x},\mathbf{w}}\max~~~\frac{\sigma_0^2|\mathbf{w}^H\overline{\mathbf{X}}\mathbf{u}_0|^2}
{\mathbf{w}^H\big[\sum_{l=-L}^L\overline{\mathbf{J}}_l\overline{\mathbf{X}}\mathbf{M}_l\overline{\mathbf{X}}^H\overline{\mathbf{J}}_l^H+\sigma_\text{r}^2\mathbf{I}\big]\mathbf{w}}\\
&~\text{s.t.}~~~~\Re\big\{\widetilde{\mathbf{h}}_i^H\mathbf{x}\big\}\geq \gamma_i,~~ \forall i = 1, \ldots, 2K_\text{u}MN, \\
&~~~~~~~~\left\|\mathbf{x}\right\|^2 = P,\\
&~~~~~~~~~\mathbf{x}\in\mathcal{X},
\end{align}
\end{subequations}
where set $\mathcal{X}$ contains the feasible solutions under certain waveform constraints (\ref{eq:CE constraint}), (\ref{eq:PAR constraint}), or (\ref{eq:similarity constraint}), not all of which are convex.
We should emphasize that since the radar system usually requires relatively high transmit power, the communication QoS requirements (\ref{eq:original problem}b) are usually satisfied and thus problem (\ref{eq:original problem}) has a  feasible solution.

It can be observed that with a fixed transmit waveform $\mathbf{x}$, the original problem (\ref{eq:original problem}) becomes a well-known minimum variance distortionless response (MVDR) problem \cite{Tang TSP 2020}, \cite{Wu TSP 2018}:
\begin{subequations}
\begin{align}
&\underset{\mathbf{w}}\min~~~\mathbf{w}^H\Big[\sum_{l=-L}^L\overline{\mathbf{J}}_l\overline{\mathbf{X}}\mathbf{M}_l
\overline{\mathbf{X}}^H\overline{\mathbf{J}}_l^H+\sigma_\text{r}^2\mathbf{I}\Big]\mathbf{w}\\
&~\text{s.t.}~~~~\mathbf{w}^H\overline{\mathbf{X}}\mathbf{u}_0 = 1.
\end{align}
\end{subequations}
The closed-form optimal solution $\mathbf{w}^\star$ in this case can be easily obtained as
\be\label{eq:optimal w}
\mathbf{w}^\star = \frac{\big[\sum_{l=-L}^L\overline{\mathbf{J}}_l\overline{\mathbf{X}}\mathbf{M}_l\overline{\mathbf{X}}^H\overline{\mathbf{J}}_l^H+\sigma_\text{r}^2\mathbf{I}\big]^{-1}\overline{\mathbf{X}}\mathbf{u}_0}
{\mathbf{u}_0^H\overline{\mathbf{X}}^H\big[\sum_{l=-L}^L\overline{\mathbf{J}}_l\overline{\mathbf{X}}\mathbf{M}_l\overline{\mathbf{X}}^H\overline{\mathbf{J}}_l^H+\sigma_\text{r}^2\mathbf{I}\big]^{-1}\overline{\mathbf{X}}\mathbf{u}_0}.
\ee
Unlike the cyclic optimization algorithms of \cite{Tang TSP 2016}-\cite{Tang TSP 2020}, we propose to directly optimize the joint transmit waveform and receive filter by substituting $\mathbf{w}^\star$ into the original optimization problem (\ref{eq:original problem}), which leads to the concentrated transmit waveform design problem:
\begin{subequations}\label{eq:problem}
\begin{align}\label{eq:problem obj}
&\underset{\mathbf{x}}\min~~-\mathbf{u}_0^H\overline{\mathbf{X}}^H\Big[\sum_{l=-L}^L\overline{\mathbf{J}}_l
\overline{\mathbf{X}}\mathbf{M}_l\overline{\mathbf{X}}^H\overline{\mathbf{J}}_l^H+\sigma_\text{r}^2\mathbf{I}\Big]^{-1}\overline{\mathbf{X}}\mathbf{u}_0\\
\label{eq:problem convex constraint}
&~\text{s.t.}~~~\Re\big\{\widetilde{\mathbf{h}}_i^H\mathbf{x}\big\}\geq \gamma_i,~~ \forall i, \\
&~~~~~~~\left\|\mathbf{x}\right\|^2 = P,\\
&~~~~~~~~\mathbf{x}\in\mathcal{X}.
\end{align}
\end{subequations}

We observe that (\ref{eq:problem}) is a complicated non-convex optimization problem due to the non-convex objective function (\ref{eq:problem obj}) and the non-convex waveform constraints (\ref{eq:problem}c), (\ref{eq:problem}d), which prevent a direct closed-form solution.
In order to tackle these difficulties, in the following sections we employ the MM and neADMM methods to covert the waveform design problems under different constraints into tractable sub-problems, and then develop efficient algorithms to iteratively solve them.

\section{Constant-Modulus Waveform Design}\label{sec:joint design}

In this section, we focus on constant-modulus waveform design for the considered STAP-SLP-based MIMO-DFRC system by developing an MM-neADMM based algorithm. As shown in the next section, this approach can be generalized to handle the other waveform constraints discussed earlier.
Substituting the constant-modulus waveform constraint (\ref{eq:CE constraint}) into (\ref{eq:problem}d), the constant-modulus waveform design problem can be formulated as
\begin{subequations}\label{eq:CE problem}
\begin{align}
&\underset{\mathbf{x}}\min~~-\mathbf{u}_0^H\overline{\mathbf{X}}^H\Big[\sum_{l=-L}^L\overline{\mathbf{J}}_l
\overline{\mathbf{X}}\mathbf{M}_l\overline{\mathbf{X}}^H\overline{\mathbf{J}}_l^H+\sigma_\text{r}^2\mathbf{I}\Big]^{-1}\overline{\mathbf{X}}\mathbf{u}_0\\
&~\text{s.t.}~~~\Re\big\{\widetilde{\mathbf{h}}_i^H\mathbf{x}\big\}\geq \gamma_i,~~ \forall i, \\
&~~~~~~~|x_j| = \sqrt{P/(MNN_\text{t})},~~\forall j,
\end{align}
\end{subequations}
where we drop the total power constraint (\ref{eq:problem}c) since the constant-modulus constraints in (\ref{eq:CE problem}c) naturally satisfy it.

\subsection{Reformulation}

In the objective function (\ref{eq:CE problem}a), the matrix $\overline{\mathbf{X}}$ contains the variable $\mathbf{x}$ to be optimized, but the relationship between them is not explicit.
Moreover, the matrix form of $\overline{\mathbf{X}}$ is not amenable for optimization.
Therefore, in order to facilitate the algorithm development, we reformulate the objective function (\ref{eq:CE problem}a) into a more favorable expression with respect to the waveform vector $\mathbf{x}$.

{\color{black}Recalling the expression for $\overline{\mathbf{X}}$ in (\ref{eq:Xbar}), we can re-write the term $\overline{\mathbf{X}}\mathbf{u}_0$ as
\begin{subequations}\label{eq:Xu0}\begin{align}
\overline{\mathbf{X}}\mathbf{u}_0 &= \left[\begin{array}{ccc}
\mathbf{I}_{N_\text{r}}\otimes \mathbf{X}_1^T & &\\
&  \ddots &\\
& & \mathbf{I}_{N_\text{r}}\otimes \mathbf{X}_M^T
\end{array}\right] \left[\begin{array}{c}\mathbf{u}_{0,1}\\ \vdots \\
\mathbf{u}_{0,M}\end{array}\right] \\
& = \left[\begin{array}{c}(\mathbf{I}_{N_\text{r}}\otimes \mathbf{X}_1^T)\mathbf{u}_{0,1}\\ \vdots \\
(\mathbf{I}_{N_\text{r}}\otimes \mathbf{X}_M^T)\mathbf{u}_{0,M}\end{array}\right],
\end{align}\end{subequations}
where $\mathbf{u}_0 = [\mathbf{u}_{0,1}^T,\ldots,\mathbf{u}_{0,M}^T]^T$ with $m$-th subvector $\mathbf{u}_{0,m}\in\mathbb{C}^{N_\text{t}N_\text{r}}$.
Using the properties of the Kronecker product \cite{Horn CUP 1990}, the $m$-th term in (\ref{eq:Xu0}b) can be re-arranged as
\begin{subequations}\label{eq:Xmu0}\begin{align}
(\mathbf{I}_{N_\text{r}}\otimes \mathbf{X}_m^T)\mathbf{u}_{0,m} & = \text{vec}\{\mathbf{X}_m^T\mathbf{U}_{0,m}\}\\
& = (\mathbf{U}_{0,m}^T\otimes\mathbf{I}_N)\text{vec}\{\mathbf{X}_m^T\},
\end{align}\end{subequations}
where the matrix $\mathbf{U}_{0,m}\in\mathbb{C}^{N_\text{t}\times N_\text{r}}$ is a reshaped version of $\mathbf{u}_{0,m}$ and $\mathbf{u}_{0,m} = \text{vec}\{\mathbf{U}_{0,m}\}$.
We can establish the following relationship between $\text{vec}\{{\mathbf{X}}_m^T\}$ and $\text{vec}\{{\mathbf{X}_m}\}$:
\be
\text{vec}\{{\mathbf{X}}_m^T\} = \mathbf{T}\text{vec}\{{\mathbf{X}_m}\},
\ee
by employing a permutation matrix $\mathbf{T}\in\mathbb{C}^{NN_\text{t}\times NN_\text{t}}$, which is defined by \cite{Hardy 2019}
\be
\mathbf{T} = \sum_{i=1}^{N_\text{t}}\sum_{j=1}^N(\mathbf{e}_{j,N}\otimes\mathbf{e}_{i,N_\text{t}})
(\mathbf{e}_{i,N_\text{t}}\otimes\mathbf{e}_{j,N})^T.
\ee
Thus, we have
\be\label{eq:Xu0a}
(\mathbf{I}_{N_\text{r}}\otimes \mathbf{X}_m^T)\mathbf{u}_{0,m}
 = \mathbf{A}_{0,m}\mathbf{x}_m,
\ee
where we define $\mathbf{A}_{0,m}\triangleq (\mathbf{U}_{0,m}^T\otimes\mathbf{I}_N)\mathbf{T}$ and $\mathbf{x}_m\triangleq\text{vec}\{{\mathbf{X}_m}\}$ for brevity.
Substituting (\ref{eq:Xu0a}) into (\ref{eq:Xu0}), the term $\overline{\mathbf{X}}\mathbf{u}_0$ becomes
\be\label{eq:A0x}
\overline{\mathbf{X}}\mathbf{u}_0 = \left[\begin{array}{c}
\mathbf{A}_{0,1}\mathbf{x}_1 \\ \vdots \\
\mathbf{A}_{0,M}\mathbf{x}_M
\end{array}\right] = \mathbf{A}_0\mathbf{x},
\ee
where we define $\mathbf{A}_0\triangleq\left[\begin{array}{ccc}
\mathbf{A}_{0,1} & & \\
& \ddots & \\
& & \mathbf{A}_{0,M}
\end{array}\right]$, and $\mathbf{x} \triangleq \text{vec}\{\mathbf{X}\} = [\mathbf{x}_1^T,\ldots,\mathbf{x}_M^T]^T$.}

\newcounter{TempEqCnt}
\setcounter{TempEqCnt}{\value{equation}}
\setcounter{equation}{39}
\begin{figure*}[!t]
\begin{subequations}\label{eq:my surrogate}
\begin{align}
f({\mathbf{x}},\mathbf{X})&\leq 2\text{Tr}\bigg\{\Big[\sum_{l=-L}^L\sum_{r=1}^{R_l}
\mathbf{A}_{l,r}\mathbf{X}_t\mathbf{A}_{l,r}^H+\sigma_\text{r}^2\mathbf{I}\Big]^{-1}
\mathbf{A}_0{\mathbf{x}}_t{\mathbf{x}}_t^H\mathbf{A}_0^H\Big[\sum_{l=-L}^L\sum_{r=1}^{R_l}
\mathbf{A}_{l,r}\mathbf{X}_t\mathbf{A}_{l,r}^H+\sigma_\text{r}^2\mathbf{I}\Big]^{-1}\Big[\sum_{l=-L}^L\sum_{r=1}^{R_l}
\mathbf{A}_{l,r}\mathbf{X}\mathbf{A}_{l,r}^H+\sigma_\text{r}^2\mathbf{I}\Big]\bigg\}\nonumber \\
&\hspace{1cm}-2\Re\Big\{{\mathbf{x}}_t^H\mathbf{A}_0^H\Big[\sum_{l=-L}^L\sum_{r=1}^{R_l}
\mathbf{A}_{l,r}\mathbf{X}_t\mathbf{A}_{l,r}^H+\sigma_\text{r}^2\mathbf{I}\Big]^{-1}
\mathbf{A}_0{\mathbf{x}}\Big\} + c_1\\
& = \text{Tr}\big\{{\mathbf{D}}_t\mathbf{X}\big\} -\Re\big\{{\mathbf{b}}_t^H\mathbf{x}\big\} + c_2,\\
& = {\mathbf{x}}^H{\mathbf{D}}_t{\mathbf{x}}-\Re\big\{{\mathbf{b}}_t^H{\mathbf{x}}\big\} + c_2.
\end{align}
\end{subequations}
\rule[-0pt]{18.5 cm}{0.05em}
\end{figure*}
\setcounter{equation}{\value{TempEqCnt}}

Since the inner CCM $\mathbf{M}_l$ is a semi-definite matrix by its definition in (\ref{eq:Ml}), it can be expressed as
\be
\mathbf{M}_l = \sum_{r=1}^{R_l}\lambda_{l,r}\widetilde{\mathbf{u}}_{l,r}\widetilde{\mathbf{u}}_{l,r}^H
= \sum_{r=1}^{R_l}\mathbf{u}_{l,r}\mathbf{u}_{l,r}^H\;,
\ee
where $R_l$ is the rank of $\mathbf{M}_l$ which generally is a small number in practice, $\lambda_{l,r}$ and $\widetilde{\mathbf{u}}_{l,r}$ are the $r$-th nonzero eigenvalue and its corresponding eigenvector, respectively, and $\mathbf{u}_{l,r} \triangleq \sqrt{\lambda_{l,r}}\widetilde{\mathbf{u}}_{l,r}$.
Hence, similar to the derivations in (\ref{eq:Xu0})-(\ref{eq:A0x}), we can re-write the term $\overline{\mathbf{J}}_l\overline{\mathbf{X}}\mathbf{M}_l\overline{\mathbf{X}}^H\overline{\mathbf{J}}_l^H$ as
\begin{subequations}\label{eq:XMX}
\begin{align}
\overline{\mathbf{J}}_l\overline{\mathbf{X}}\mathbf{M}_l\overline{\mathbf{X}}^H\overline{\mathbf{J}}_l^H &=
\sum_{r=1}^{R_l}\overline{\mathbf{J}}_l\overline{\mathbf{X}}\mathbf{u}_{l,r}\mathbf{u}_{l,r}^H\overline{\mathbf{X}}^H\overline{\mathbf{J}}_l^H \\
&= \sum_{r=1}^{R_l}\mathbf{A}_{l,r}{\mathbf{x}}{\mathbf{x}}^H\mathbf{A}_{l,r}^H
\end{align}
\end{subequations}
\be
\mathbf{A}_{l,r} \triangleq \overline{\mathbf{J}}_l\left[\begin{array}{ccc} \mathbf{A}_{l,r,1} & & \\
& \ddots & \\ & & \mathbf{A}_{l,r,M} \end{array}\right] \; ,
\ee
where $\mathbf{A}_{l,r,m} \triangleq (\mathbf{U}_{l,r,m}^T\otimes\mathbf{I}_N)\mathbf{T}$ and $\mathbf{U}_{l,r,m}\in\mathbb{C}^{N_\text{t}\times N_\text{r}}$ is a reshaped version of the $m$-th sub-vector of $\mathbf{u}_{l,r}$, namely $\mathbf{u}_{l,r,m}\in\mathbb{C}^{N_\text{t}N_\text{r}}$. Based on the results in (\ref{eq:A0x}) and (\ref{eq:XMX}b), the objective function (\ref{eq:CE problem}a) can be equivalently re-formulated as \setcounter{equation}{40}
\be\label{eq:reformulate obj}
-{\mathbf{x}}^H\mathbf{A}_0^H\Big[\sum_{l=-L}^L\sum_{r=1}^{R_l}
\mathbf{A}_{l,r}{\mathbf{x}}{\mathbf{x}}^H\mathbf{A}_{l,r}^H
+\sigma_\text{r}^2\mathbf{I}\Big]^{-1}\mathbf{A}_0{\mathbf{x}}.
\ee

\subsection{MM Transformation}

\textcolor{black}{In order to efficiently solve the waveform design problem, we first utilize the MM method to convert it into a sequence of simpler problems to be solved until convergence.
Specifically, given the obtained solution $\mathbf{x}_t$ in the $t$-th iteration, we attempt to construct a more tractable surrogate function that approximates the complicated non-convex objective function (\ref{eq:reformulate obj}) at the current local point $\mathbf{x}_t$ and serves as an upper-bound to be minimized in the next iteration.}
The following lemma \cite{Sun TSP 17} is utilized to find a surrogate function for (\ref{eq:reformulate obj}).
\begin{lemma} \label{lemma1}
For a positive-definite matrix $\mathbf{W}$, the function $-\mathbf{s}^H\mathbf{W}^{-1}\mathbf{s}$ is concave in  $\mathbf{s}$ and $\mathbf{W}$, and is therefore upper-bounded by its first-order linear expansion around $(\mathbf{s}_t,\mathbf{W}_t)$ as
\be\label{eq:surrogate function}
- \mathbf{s}^H\mathbf{W}^{-1}\mathbf{s} \leq  2\text{Tr}\big\{\mathbf{W}_\text{t}^{-1}
\mathbf{s}_\text{t}\mathbf{s}_\text{t}^H\mathbf{W}_\text{t}^{-1}\mathbf{W}\big\} - 2\Re\big\{\mathbf{s}_\text{t}^H\mathbf{W}_\text{t}^{-1}\mathbf{s}\big\} + c, \nonumber
\ee
\hspace{-0.1 cm} where $c$ is a constant term that is irrelevant to the variables.

\hfill $\blacksquare$
\end{lemma}%

In order to utilize the findings in Lemma 1, we define following notation:
\begin{subequations}\begin{align}
\mathbf{s} &\triangleq \mathbf{A}_0{\mathbf{x}},\\
\mathbf{X} &\triangleq {\mathbf{x}}{\mathbf{x}}^H,\\
\mathbf{W} &\triangleq \sum_{l=-L}^L\sum_{r=1}^{R_l}
\mathbf{A}_{l,r}\mathbf{X}\mathbf{A}_{l,r}^H
+\sigma_\text{r}^2\mathbf{I},
\end{align}\end{subequations}
and we write the objective function in~(\ref{eq:reformulate obj}) as $f({\mathbf{x}},\mathbf{X})$.
Then, the surrogate function of $f({\mathbf{x}},\mathbf{X})$ at point $({\mathbf{x}}_t,\mathbf{X}_t)$, \textcolor{black}{where $\mathbf{x}_t$ is the obtained solution in the $t$-th iteration and $\mathbf{X}_t \triangleq \mathbf{x}_t\mathbf{x}_t^H$,} can be calculated as in~(\ref{eq:my surrogate}) presented at the top of this page, where for brevity we define
\begin{subequations}\begin{align}
{\mathbf{b}}_t &\triangleq 2\mathbf{A}_0^H\Big[\sum_{l=-L}^L\sum_{r=1}^{R_l}
\mathbf{A}_{l,r}\mathbf{X}_t\mathbf{A}_{l,r}^H
+\sigma_\text{r}^2\mathbf{I}\Big]^{-1}\mathbf{A}_0\mathbf{x}_t, \\
{\mathbf{D}}_t &\triangleq 2\sum_{l=-L}^L\sum_{r=1}^{R_l}\mathbf{G}_{t,l,r}^H\mathbf{X}_t\mathbf{G}_{t,l,r},\\
\mathbf{G}_{t,l,r} &\triangleq \mathbf{A}_0^H\Big[\sum_{l=-L}^L\sum_{r=1}^{R_l}
\mathbf{A}_{l,r}\mathbf{X}_t\mathbf{A}_{l,r}^H
+\sigma_\text{r}^2\mathbf{I}\Big]^{-1}\mathbf{A}_{l,r}.
\end{align}\end{subequations}
The constant terms $c_1$ and $c_2$ are irrelevant to the variables ${\mathbf{x}}$ and $\mathbf{X}$, and thus their detailed expressions are omitted.
Equation (\ref{eq:my surrogate}c) is obtained by substituting $\mathbf{X} \triangleq {\mathbf{x}}{\mathbf{x}}^H$ back into (\ref{eq:my surrogate}b).

Based on above derivations, the transmit waveform design problem at point $\mathbf{x}_t$ can be formulated as
\begin{subequations}\label{eq:surrogate problem}
\begin{align}\label{eq:surrogate problem a}
&\underset{\mathbf{x}}\min~~~\mathbf{x}^H\mathbf{D}_t\mathbf{x}-\Re\big\{\mathbf{b}_t^H\mathbf{x}\big\}\\
&~\text{s.t.}~~~~\Re\big\{\widetilde{\mathbf{h}}_i^H\mathbf{x}\big\}\geq \gamma_i,~~ \forall i, \\
\label{eq:nonlinear equality constraint}
&~~~~~~~~\big|x_j\big| = \sqrt{P/(MNN_\text{t})},~~\forall j.
\end{align}
\end{subequations}
It can be observed that although the objective function (\ref{eq:surrogate problem a}) is continuous and convex, problem (\ref{eq:surrogate problem}) is still a non-convex problem due to the constant-modulus constraint (\ref{eq:nonlinear equality constraint}).
\textcolor{black}{While relaxing the non-convex equality constraint (\ref{eq:nonlinear equality constraint}) is an obvious approach, solving the problem with the relaxed constraint and then projecting the solution onto the constraint leads to a significant performance loss, so we propose to directly cope with the equality constraint by employing the neADMM method. }
While the classical ADMM method can only handle linear equality constraints, the new neADMM approach \cite{Liu TCOM 2021} can be applied to nonlinear equality constraints such as~(\ref{eq:nonlinear equality constraint}).
Therefore, we develop an neADMM-based method to solve this problem as follows.

\subsection{neADMM Transformation}

We first introduce an auxiliary variable $\mathbf{y}\triangleq [y_1,\ldots,y_{MNN_\text{t}}]^T$ to decouple the convex constraint (\ref{eq:surrogate problem}b) and the non-convex constraint (\ref{eq:surrogate problem}c) with respect to $\mathbf{x}$, and convert problem (\ref{eq:surrogate problem}) to
\begin{subequations}\label{eq:x y problem}
\begin{align}
&\underset{\mathbf{x},\mathbf{y}}\min~~~\mathbf{x}^H\mathbf{D}_t\mathbf{x}-\Re\big\{\mathbf{b}_t^H\mathbf{x}\big\}\\
\label{eq:x y problem c1}
&~\text{s.t.}~~~~\Re\big\{\widetilde{\mathbf{h}}_i^H\mathbf{x}\big\}\geq \gamma_i,~~ \forall i, \\
\label{eq:x y problem c2}
&~~~~~~~~\big|x_j\big| \leq \sqrt{P/(MNN_\text{t})},~~\forall j,\\
\label{eq:linear equality}
&~~~~~~~~\mathbf{x} = \mathbf{y},\\
\label{eq:nonlinear equality}
&~~~~~~~~\big|y_j\big| = \sqrt{P/(MNN_\text{t})},~~\forall j.
\end{align}
\end{subequations}
To accommodate the neADMM framework, we define the feasible region of the inequality constraints (\ref{eq:x y problem c1}) and (\ref{eq:x y problem c2}) as set $\mathcal{C}$, and an indicator function $\mathbb{I}_\mathcal{C}$ associated with the set $\mathcal{C}$ as
\begin{equation}\label{eq:indicator function}
\mathbb{I}_\mathcal{C}(\mathbf{x}) = \left\{
             \begin{array}{lr}0,\hspace{0.75 cm}\mathbf{x}\in\mathcal{C},\\
             +\infty,\hspace{0.3 cm}\text{otherwise}.
             \end{array}
\right.
\end{equation}
Then, by removing the constraints on $\mathbf{x}$ and adding the feasibility indicator function in  the objective, problem~(\ref{eq:x y problem}) is transformed to
\begin{subequations}\label{eq:xy indicator problem}
\begin{align}
&\underset{\mathbf{x},\mathbf{y}}\min~~~\mathbf{x}^H\mathbf{D}_t\mathbf{x}
-\Re\big\{\mathbf{b}_t^H\mathbf{x}\big\} + \mathbb{I}_\mathcal{C}(\mathbf{x})\\
&~\text{s.t.}~~~~\mathbf{x} = \mathbf{y},\\
&~~~~~~~~\big|y_j\big| = \sqrt{P/(MNN_\text{t})},~~\forall j,
\end{align}
\end{subequations}
whose solution can be obtained by optimizing its augmented Lagrangian (AL) function.
Specifically, the AL function of problem (\ref{eq:xy indicator problem}) is expressed as
\be\label{eq:AL function}\begin{aligned}
&\mathcal{L}(\mathbf{x},\mathbf{y},\bm{\lambda},\bm{\mu}) \triangleq \mathbf{x}^H\mathbf{D}_t\mathbf{x}-\Re\big\{\mathbf{b}_t^H\mathbf{x}\big\} +\mathbb{I}_\mathcal{C}(\mathbf{x})\\&~+ \frac{\rho}{2}\big\|\mathbf{x}\hspace{-0.1 cm}-\hspace{-0.1 cm}\mathbf{y}\hspace{-0.1 cm}+\hspace{-0.1 cm}\bm{\lambda}/\rho\big\|^2 + \frac{\rho}{2}\big\||\mathbf{y}|\hspace{-0.1 cm}-\hspace{-0.1 cm}\sqrt{P/(MNN_\text{t})}\hspace{-0.1 cm}+\hspace{-0.1 cm}\bm{\mu}/\rho\big\|^2,
\end{aligned}\ee
where $\rho>0$ is a penalty parameter, $\bm{\lambda}\in\mathbb{C}^{MNN_\text{t}}$ and  $\bm{\mu}\in\mathbb{C}^{MNN_\text{t}}$ are dual variables, and $|\cdot|$ is an element-wise absolute value operation. The AL function (\ref{eq:AL function}) is a more tractable function with multiple variables, which can be minimized by alternately updating $\mathbf{x}$, $\mathbf{y}$, $\bm{\lambda}$, and $\bm{\mu}$ as shown below.

\subsection{Block Update}

1) \textbf{\textit{Update $\mathbf{x}$}}: With $\mathbf{y}$, $\bm{\lambda}$ and $\bm{\mu}$ given, the optimization problem for updating $\mathbf{x}$ is formulated as
\be\label{eq:temp problem x}
\underset{\mathbf{x}}\min~~~\mathbf{x}^H\mathbf{D}_t\mathbf{x}-\Re\big\{\mathbf{b}_t^H\mathbf{x}\big\} +\mathbb{I}_\mathcal{C}(\mathbf{x})+ \frac{\rho}{2}\big\|\mathbf{x}-\mathbf{y}+\bm{\lambda}/\rho\big\|^2.
\ee
According to the definition of $\mathbb{I}_\mathcal{C}(\mathbf{x})$ in (\ref{eq:indicator function}), problem (\ref{eq:temp problem x}) can be equivalently transformed into a convex second-order cone programming (SOCP) problem:
\begin{subequations}\label{eq:problem x}
\begin{align}
&\underset{\mathbf{x}}\min~~~\mathbf{x}^H\mathbf{D}_t\mathbf{x}-\Re\big\{\mathbf{b}_t^H\mathbf{x}\big\} + \frac{\rho}{2}\big\|\mathbf{x}-\mathbf{y}+\bm{\lambda}/\rho\big\|^2\\
&~\text{s.t.}~~~~\Re\big\{\widetilde{\mathbf{h}}_i^H\mathbf{x}\big\}\geq \gamma_i,~~ \forall i, \\
&~~~~~~~~\big|x_j\big| \leq \sqrt{P/(MNN_\text{t})},~~\forall j.
\end{align}
\end{subequations}
\textcolor{black}{whose optimal solution $\mathbf{x}^\star$ can be readily obtained by various off-the-shelf algorithms and optimization tools such as those proposed in \cite{SOCP1}, \cite{SOCP2}. In addition, the algorithm \cite{Liu JSTSP 2021} that employs the Lagrangian dual with the aid of the Hooke-Jeeves Pattern Search method can be utilized to offer an efficient solution.}

2) \textbf{\textit{Update $\mathbf{y}$}}:
With fixed $\mathbf{x}$, $\bm{\lambda}$ and $\bm{\mu}$, the optimization problem for updating $\mathbf{y}$ is given by
\be\label{eq:problem y}
\underset{\mathbf{y}}\min~~\frac{\rho}{2}\big\|\mathbf{x}-\mathbf{y}+\bm{\lambda}/\rho\big\|^2+ \frac{\rho}{2}\big\||\mathbf{y}|-\sqrt{P/(MNN_\text{t})}+\bm{\mu}/\rho\big\|^2.
\ee
We observe that problem (\ref{eq:problem y}) is a non-convex problem due to the absolute value operation.
Fortunately, problem~(\ref{eq:problem y}) is separable in the elements of $\mathbf{y}$, and
thus we can equivalently divide~(\ref{eq:problem y}) into $MNN_\text{t}$ sub-problems.
The $i$-th sub-problem is expressed as
\be\label{eq:problem ym}
\underset{y_i}\min~~~\big|y_i-a_i\big|^2+ \big||y_i|-b_i\big|^2,
\ee
where $a_i$ and $b_i$ are the $i$-th element of $\mathbf{x}+\bm{\lambda}/\rho$ and $\sqrt{P/(MNN_\text{t})}-\bm{\mu}/\rho$, respectively.
In order to handle the absolute value function, the objective of (\ref{eq:problem ym}) is expanded as
\begin{subequations}\begin{align}
&\hspace{-0.2 cm} \big|y_i-a_i\big|^2+ \big||y_i|-b_i\big|^2\\
&= 2|y_i|^2-2\Re\big\{(a_i^*y_i+b_i^*|y_i|)\big\}\hspace{-0.04cm}+\hspace{-0.04cm}|a_i|^2\hspace{-0.04cm}+\hspace{-0.04cm}|b_i|^2\\
\label{eq:ym abs}
&= 2|y_i|^2\hspace{-0.06cm}-\hspace{-0.06cm}2|y_i|\Re\big\{(a_i^*e^{\jmath\angle y_i}\hspace{-0.08cm}+\hspace{-0.06cm}b_i^*)\big\}\hspace{-0.07cm}+\hspace{-0.07cm}|a_i|^2
\hspace{-0.07cm}+\hspace{-0.07cm}|b_i|^2.
\end{align}\end{subequations}
Since $|y_i|\geq0$, we can easily obtain the optimal angle of $y_i$ as $
\angle y_i^\star = \angle a_i$.
Substituting $\angle y_i^\star$ into (\ref{eq:ym abs}), the optimal amplitude of $y_i$ can be obtained by solving
\be
\underset{|y_i|}\min~~~2|y_i|^2-2|y_i|(|a_i|+\Re\{b_i\}),
\ee
whose optimal solution is given by $|y_i^\star| = 0.5(|a_i|+\Re\{b_i\})$.
Therefore, the optimal solution to problem (\ref{eq:problem ym}) is
\be\label{eq:optimal ym}
y_i^\star = 0.5\big(|a_i|+\Re\{b_i\}\big)e^{\jmath\angle a_i}.
\ee

3) \textbf{\textit{Update $\bm{\lambda}$ and $\bm{\mu}$}}:
After obtaining $\mathbf{x}$ and $\mathbf{y}$, the dual variables $\bm{\lambda}$ and $\bm{\mu}$ are updated by
\begin{subequations}\label{eq:update dual variables}
\begin{align}
\label{eq:update lambda}
\bm{\lambda}^\star &:= \bm{\lambda} + \rho(\mathbf{x}-\mathbf{y}),\\
\label{eq:update mu}
\bm{\mu}^\star &:= \bm{\mu} + \rho\big[|\mathbf{y}|-\sqrt{P/(MNN_\text{t})}\big].
\end{align}\end{subequations}

\begin{algorithm}[!t]
\begin{small}
\caption{Proposed MM-neADMM Algorithm for Constant-Modulus Waveform Design}
\label{alg1}
    \begin{algorithmic}[1]
    \REQUIRE $\mathbf{A}_0$, $\mathbf{A}_{l,r}$, $\forall l$, $\forall r$, $\widetilde{\mathbf{h}}_i$, $\gamma_i$, $\forall i$, $P$, $\rho$.
    \ENSURE $\mathbf{x}^\star$.
        \STATE {Initialize $\mathbf{x}$ by solving (\ref{eq:initialization problem}), $\mathbf{y}:=\mathbf{x}$, $\bm{\lambda}:=\mathbf{0}$, $\bm{\mu}:=\mathbf{0}$.}
        \WHILE {no convergence }
            \STATE{Update $\mathbf{x}$ by solving (\ref{eq:problem x}).}
            \STATE{Update $y_i,~\forall i$, by (\ref{eq:optimal ym}).}
            \STATE{Update $\bm{\lambda}$ by (\ref{eq:update lambda}).}
            \STATE{Update $\bm{\mu}$ by (\ref{eq:update mu}).}
        \ENDWHILE
        \STATE{$\mathbf{x}^\star = \mathbf{x}$.}
    \end{algorithmic}
    \end{small}
\end{algorithm}

\subsection{Summary, Initialization, and Complexity Analysis}

With the above derivations, the proposed MM-neADMM algorithm for constant-modulus waveform design is straightforward and summarized in Algorithm 1.
In summary, the transmit waveform $\mathbf{x}$ is obtained by iteratively updating $\mathbf{x}$, $\mathbf{y}$, $\bm{\lambda}$ and $\bm{\mu}$ via (\ref{eq:problem x}), (\ref{eq:optimal ym}), (\ref{eq:update lambda}) and (\ref{eq:update mu}), respectively, until the relative increase of the achieved radar output SINR is less than a given convergence threshold.
Finally, with the obtained transmit waveform $\mathbf{x}^\star$, the optimal receive filter $\mathbf{w}^\star$ can be calculated by (\ref{eq:optimal w}).

Since a good starting point is preferable for the proposed alternating optimization algorithm, we investigate how to properly initialize $\mathbf{x}$ before the iterations.
In order to retain DoFs to maximize the radar output SINR under the given communication QoS constraints, we propose to use the intuitive approach of initializing $\mathbf{x}$ by maximizing the minimum QoS of the communication users using the available transmit power.
Therefore, the optimization problem for initialization is formulated as
\begin{subequations}\label{eq:initialization problem}
\begin{align}
\underset{\mathbf{x}}\max~~&\underset{i}\min~~\Re\big\{\widetilde{\mathbf{h}}_i^H\mathbf{x}\big\}\\
&~\text{s.t.}~~~\big|x_j\big| \leq \sqrt{P/(MNN_\text{t})},~~\forall j,
\end{align}
\end{subequations}
where the power constraint (\ref{eq:initialization problem}b) is a relaxed convex version of the constant-modulus constraint (\ref{eq:CE constraint}) for the purpose of simplifying the solution.
It is obvious that problem (\ref{eq:initialization problem}) is convex and can be efficiently solved by the interior point method, CVX, etc.

Next, we briefly analyze the computational complexity of the proposed waveform design algorithm.
We assume that the typical interior point method is employed to solve the SOCP problem (\ref{eq:problem x}).
Since problem (\ref{eq:problem x}) has an $MNN_\text{t}$-dimensional variable with $2K_\text{u}MN$ linear matrix inequality (LMI) constraints and $MNN_\text{t}$ second-order cone (SOC) constraints, the computational complexity to update $\mathbf{x}$ is of order $\mathcal{O}\{\ln(1/\varpi)\sqrt{(4K_\text{u}+N_\text{t})MN}M^2N^2N_\text{t}(2K_\text{u}+N_\text{t} +2MNN^2_\text{t}\}$ with $\varpi$ representing the convergence threshold.
The closed-form update for $\mathbf{y}$ in (\ref{eq:optimal ym}) or for $\bm{\lambda}$ and $\bm{\mu}$ in (\ref{eq:update dual variables}) requires the same order of computational complexity $\mathcal{O}\{MNN_\text{t}\}$.
\textcolor{black}{Thus, the total computational complexity to solve for the waveform vector mainly depends on the update for $\mathbf{x}$, which emphasizes the need for low-complexity algorithms that handle large-scale SOCP problems.}

\section{Generalizations to Other Waveform Constraints}
\vspace{0.0 cm}

In this section, we generalize the proposed MM-neADMM algorithm when other waveform constraints such as the PAPR constraint (\ref{eq:PAR constraint}) or the similarity constraint (\ref{eq:similarity constraint}) are employed.

\subsection{PAPR-Constrained Waveform Design}

Substituting the PAPR constraint (\ref{eq:PAR constraint}) into (\ref{eq:problem}d), the waveform design problem (\ref{eq:problem}) is re-formulated as
\begin{subequations}\label{eq:PAR problem}
\begin{align}
&\underset{\mathbf{x}}\min~~-\mathbf{u}_0^H\overline{\mathbf{X}}^H\Big[\sum_{l=-L}^L\overline{\mathbf{J}}_l
\overline{\mathbf{X}}\mathbf{M}_l\overline{\mathbf{X}}^H\overline{\mathbf{J}}_l^H+\sigma_\text{r}^2\mathbf{I}\Big]^{-1}\overline{\mathbf{X}}\mathbf{u}_0\\
&~\text{s.t.}~~~\Re\big\{\widetilde{\mathbf{h}}_i^H\mathbf{x}\big\}\geq \gamma_i,~~ \forall i, \\
&~~~~~~~\big|x_j\big| \leq \sqrt{(1+\varepsilon) P/(MNN_\text{t})},~~\forall j,\\
&~~~~~~~~\left\|\mathbf{x}\right\|^2 = P.
\end{align}
\end{subequations}
We can observe that problem (\ref{eq:PAR problem}) is similar to the constant-modulus waveform design (\ref{eq:CE problem}) except that all elements of $\mathbf{x}$ are jointly constrained in the equality constraint (\ref{eq:PAR problem}d), which requires some modifications to the proposed MM-neADMM algorithm framework as described below.

First, following the derivations in the previous section, we replace the objective (\ref{eq:PAR problem}a) with its surrogate function and introduce an auxiliary variable $\mathbf{y}$ to decouple the convex constraints (\ref{eq:PAR problem}b), (\ref{eq:PAR problem}c), and the non-convex total power constraint (\ref{eq:PAR problem}d):
\begin{subequations}\label{eq:PAR problem au}
\begin{align}
&\underset{\mathbf{x}}\min~~~\mathbf{x}^H\mathbf{D}_t\mathbf{x}-\Re\big\{\mathbf{b}_t^H\mathbf{x}\big\}\\
&~\text{s.t.}~~~~\Re\big\{\widetilde{\mathbf{h}}_i^H\mathbf{x}\big\}\geq \gamma_i,~~ \forall i, \\
&~~~~~~~~\big|x_j\big| \leq \sqrt{(1+\varepsilon) P/(MNN_\text{t})},~~\forall j,\\
&~~~~~~~~\left\|\mathbf{x}\right\|^2 \leq P,\\
&~~~~~~~~\left\|\mathbf{y}\right\|^2 = P,\\
&~~~~~~~~~\mathbf{x} = \mathbf{y}.
\end{align}
\end{subequations}
Then, defining the feasible region of the constraints (\ref{eq:PAR problem au}b)-(\ref{eq:PAR problem au}e) as set $\mathcal{E}$ and the associated indicator function $\mathbb{I}_\mathcal{E}(\mathbf{x})$, the AL function of problem (\ref{eq:PAR problem au}) can be expressed as
\be\label{eq:PAR AL problem}\begin{aligned}
\mathcal{L}(\mathbf{x},\mathbf{y},\bm{\lambda}) \triangleq & \,\, \mathbf{x}^H\mathbf{D}_t\mathbf{x}-\Re\big\{\mathbf{b}_t^H\mathbf{x}\big\} +\mathbb{I}_\mathcal{E}(\mathbf{x})\\
& \,\, + \frac{\rho}{2}\big\|\mathbf{x}-\mathbf{y}+\bm{\lambda}/\rho\big\|^2.
\end{aligned}\ee

We again propose to iteratively update each variable. Based on (\ref{eq:PAR AL problem}), the variable $\mathbf{x}$ is updated by optimizing
\begin{subequations}\label{eq:PAR update x}
\begin{align}
&\underset{\mathbf{x}}\min~~~\mathbf{x}^H\mathbf{D}_t\mathbf{x}-\Re\big\{\mathbf{b}_t^H\mathbf{x}\big\} + \frac{\rho}{2}\big\|\mathbf{x}-\mathbf{y}+\bm{\lambda}/\rho\big\|^2\\
&~\text{s.t.}~~~~\Re\big\{\widetilde{\mathbf{h}}_i^H\mathbf{x}\big\}\geq \gamma_i,~~ \forall i, \\
&~~~~~~~~\big|x_j\big| \leq \sqrt{(1+\varepsilon) P/(MNN_\text{t})},~~\forall j,\\
&~~~~~~~~\left\|\mathbf{x}\right\|^2 \leq P,
\end{align}
\end{subequations}
which is also a convex SOCP problem and can be easily solved using various existing algorithms.
The optimization problem for updating the auxiliary variable $\mathbf{y}$ is formulated as
\begin{subequations}
\begin{align}
&\underset{\mathbf{y}}\min~~~\frac{\rho}{2}\big\|\mathbf{x}-\mathbf{y}+\bm{\lambda}/\rho\big\|^2\\
&~\text{s.t.}~~~~\left\|\mathbf{y}\right\|^2 = P,
\end{align}
\end{subequations}
whose optimal solution is given by
\be\label{eq:PAR update y}
\mathbf{y}^\star = \frac{\sqrt{P}(\mathbf{x}+\bm{\lambda}/\rho)}{\left\|\mathbf{x}+\bm{\lambda}/\rho\right\|}.
\ee
Finally, the dual variable $\bm{\lambda}$ is updated by (\ref{eq:update lambda}).

\begin{algorithm}[!t]
\begin{small}
\caption{Proposed MM-neADMM Algorithm for PAPR-Constrained Waveform Design}
\label{alg2}
    \begin{algorithmic}[1]
    \REQUIRE $\mathbf{A}_0$, $\mathbf{A}_{l,r}$, $\forall l$, $\forall r$, $\widetilde{\mathbf{h}}_i$, $\gamma_i$, $\forall i$, $P$, $\rho$.
    \ENSURE $\mathbf{x}^\star$.
        \STATE {Initialize $\mathbf{x}$, $\mathbf{y}:=\mathbf{x}$, $\bm{\lambda}:=\mathbf{0}$.}
        \WHILE {no convergence}
            \STATE{Update $\mathbf{x}$ by solving (\ref{eq:PAR update x}).}
            \STATE{Update $\mathbf{y}$ by (\ref{eq:PAR update y}).}
            \STATE{Update $\bm{\lambda}$ by (\ref{eq:update lambda}).}
        \ENDWHILE
        \STATE{$\mathbf{x}^\star = \mathbf{x}$.}
    \end{algorithmic}
    \end{small}
\end{algorithm}

Given the above derivations, the proposed MM-neADMM algorithm for the PAPR-constrained waveform design is straightforward and summarized in Algorithm 2.
The initialization is obtained by solving a convex problem that has the PAPR constraint (\ref{eq:PAR problem au}c) and the  total power constraint (\ref{eq:PAR problem au}d) similar to (\ref{eq:initialization problem}).
In each iteration, the variable $\mathbf{x}$ is updated by solving an $MNN_\text{t}$-dimensional SOCP problem with $2K_\text{u}MN$ LMI constraints and $(MNN_\text{t}+1)$ SOC constraints, whose computational complexity is of order
$\mathcal{O}\{\ln(1/\varpi)\sqrt{(4K_\text{u}+N_\text{t})MN+1}MNN_\text{t}(2MNN_\text{t}(MNN_\text{t}+1)+2K_\text{u}MN)\}$.
The computational complexity of the closed-form update for $\mathbf{y}$ and $\bm{\lambda}$ are of the same order $\mathcal{O}\{MNN_\text{t}\}$, which is much lower than that of updating $\mathbf{x}$.

\subsection{Constant-Modulus and Similarity-Constrained Waveform Design}

Here we investigate the waveform design taking into account both the constant-modulus constraint and the similarity constraint between the designed waveform and a given reference.
Substituting the constant-modulus constraint (\ref{eq:CE constraint}) and similarity constraint (\ref{eq:similarity constraint}) into (\ref{eq:problem}d), the waveform design problem becomes
\begin{subequations}\label{eq:CES problem}
\begin{align}
&\underset{\mathbf{x}}\min~-\mathbf{u}_0^H\overline{\mathbf{X}}^H\Big[\sum_{l=-L}^L\overline{\mathbf{J}}_l
\overline{\mathbf{X}}\mathbf{M}_l\overline{\mathbf{X}}^H\overline{\mathbf{J}}_l^H+\sigma_\text{r}^2\mathbf{I}\Big]^{-1}\overline{\mathbf{X}}\mathbf{u}_0\\
&~\text{s.t.}~~~~\Re\big\{\widetilde{\mathbf{h}}_i^H\mathbf{x}\big\}\geq \gamma_i,~~ \forall i, \\
&~~~~~~~~|x_j| = \sqrt{P/(MNN_\text{t})},~~\forall j,\\
&~~~~~~~~|x_j - x_{0,j}| \leq \xi,~~\forall j.
\end{align}
\end{subequations}
We observe that the optimization problem (\ref{eq:CES problem}) is very similar to the constant-modulus waveform design problem (\ref{eq:CE problem}), except for the additional tractable convex similarity constraints (\ref{eq:CES problem}d).
Thus, following the procedure in Sec. III, we first re-formulate the objective function in a vector form and derive its convex surrogate function.
Then, an auxiliary variable $\mathbf{y}$ is introduced to decouple the convex constraints (\ref{eq:CES problem}b), (\ref{eq:CES problem}d) and the non-convex equality constraint (\ref{eq:CES problem}c) with respect to $\mathbf{x}$, as in (\ref{eq:x y problem}).
The neADMM algorithm is finally employed to iteratively update $\mathbf{x}$, $\mathbf{y}$, and the dual variables.
Compared with (\ref{eq:problem x}), the solution for the update to $\mathbf{x}$ must consider the additional convex similarity constraint in~(\ref{eq:CES problem}d) on each element in $\mathbf{x}$.
This results in an $MNN_\text{t}$-dimensional SOCP problem with $2K_\text{u}MN$ LMI constraints and $2MNN_\text{t}$ SOC constraints, whose computational complexity is of order $\mathcal{O}\{\ln(1/\varpi)\sqrt{2MN(2K_\text{u}+N_\text{t})}M^2N^2N_\text{t}(3MNN_\text{t}^2+2N_\text{t}+2K_\text{u}\}$.
The details are omitted here due to space limitations.

\section{Simulation Results}\label{sec:simulation results}

In this section, we provide simulation results to show the effectiveness of the proposed joint transmit waveform and receive filter design algorithms.
The following settings are assumed throughout our simulations.
The BS is equipped with the same number of transmit and receive antennas $N_\text{t} = N_\text{r} = 6 $ with antenna spacing $d_\text{t} = 2\lambda$ and $d_\text{r} = \lambda/2$, respectively.
A CPI has $M = 4$ pulses with the PRF $f_\text{r} = 1000$Hz, and each pulse is sampled $N = 8$ times.
The carrier frequency of the transmit waveform is $f_0 = 2.4$GHz and the noise power of the echoes is $\sigma_\text{r}^2 = 0$dB.
The target of interest is at the azimuth $\theta_0 = 0^\circ$ with a normalized Doppler frequency $f_\text{d}=0.3$ and power $\sigma_0^2 = 0$dB unless otherwise stated.
The clutter is assumed to be returned from the CUT and the nearest $4$ adjacent range cells with power $\sigma_\text{c}^2 = 0$dB, each of which consists of $N_\text{c} = 60$ clutter patches evenly distributed in azimuth.
The BS also transmits information symbols to $K_\text{u} = 3$ communication users, and the communication noise power is set as $\sigma^2 = -20\text{dB}$.
The communication QoS for all $K_\text{u}$ users is the same and is denoted by $\Gamma$.
The penalty parameter is set as $\rho = 1$.
\textcolor{black}{Typical orthogonal linear frequency modulated (LFM) waveforms \cite{Li SPM 2007}, \cite{Cui TSP 2017}, \cite{Wu TSP 2018} are chosen as the reference waveforms since they achieve good pulse compression and ambiguity function properties.}
The samples of the LFM waveforms are denoted by $\mathbf{X}_0\in\mathbb{C}^{N_\text{t}\times MN}$, each element of which is given by
\be\nonumber
\mathbf{X}_0(i,j) = \sqrt{\frac{P}{MNN_\text{t}}}\exp\{\jmath2\pi i(j-1)/N_\text{t}\}\exp\{\jmath\pi(j-1)^2/N_\text{t}\},
\ee
and the reference waveform vector is $\mathbf{x}_0 = \text{vec}\{\mathbf{X}_0\}$.
In the following, the proposed DFRC waveform designs under the constant-modulus constraint, the combined constant-modulus and similarity constraints, and the PAPR constraint are referred to as ``Proposed, CM'', ``Proposed, CMS'', and ``Proposed, PAPR'', respectively.
\textcolor{black}{For comparison, the STAP-based MIMO radar-only schemes under these constraints are also included and referred to as ``Radar, CM'', ``Radar, CMS'', and ``Radar, PAPR'', respectively.
In order to illustrate the advantages of the proposed CI-based SLP approach in DFRC systems, we also include the results for a zero-forcing (ZF) approach that implements the optimization of (\ref{eq:original problem}) with an equality constraint for the communication QoS constraint (\ref{eq:original problem}b). This approach, which we refer to as ``Non-CI ZF'', eliminates the MUI at the receivers but does not exploit it.}

\begin{figure}[!t]
\centering
\includegraphics[width = 3.5 in]{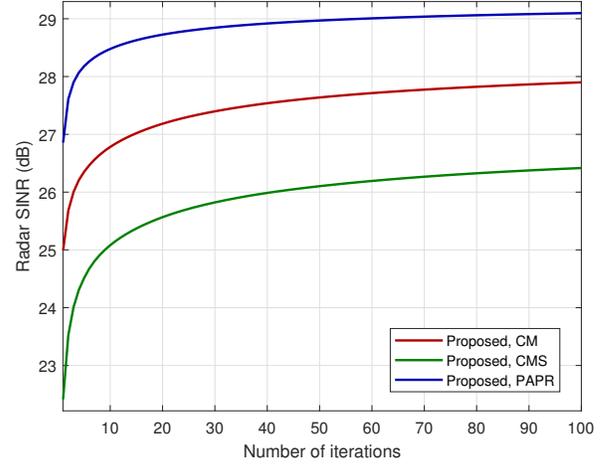}\vspace{0.1 cm}
\caption{Convergence illustration ($\Gamma = 5$dB, $\theta_0 = 0$, $f_\text{d} = 0.3$, $P = 30$W, $\xi=1.5\sqrt{\frac{P}{MNN_\text{t}}}$, $\varepsilon=1$).}\label{fig:iterations}
\end{figure}

We first show the convergence performance of the proposed algorithms in Fig. \ref{fig:iterations}, where the communication QoS is set as $\Gamma = 5$dB, the total transmit power is $P = 30$W, the similarity threshold is $\xi=1.5\sqrt{\frac{P}{MNN_\text{t}}}$, and the PAPR threshold is $\varepsilon=1$. We see that the stricter the waveform constraint, the slower the algorithm convergence, with the PAPR constraint resulting in the fastest convergence and highest SINR, while the CMS constraint yields the slowest convergence and lowest SINR. Moreover, we see that the radar SINR monotonically increases with the iterations, which is consistent with the behavior of the MM method.

\begin{figure}[!t]
\centering
\includegraphics[width = 3.5 in]{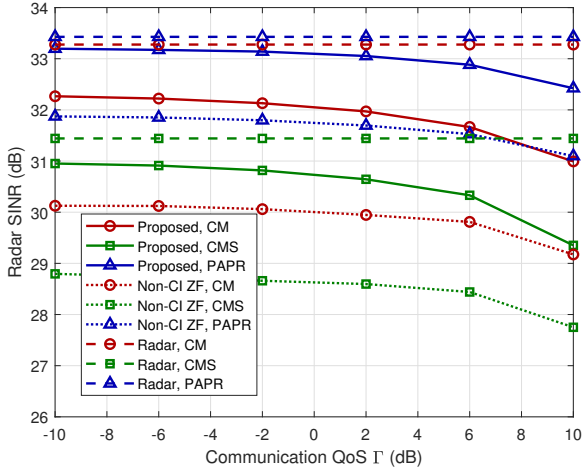}\vspace{0.1 cm}
\caption{Radar output SINR versus communication QoS ($\theta_0 = 0$, $f_\text{d} = 0.3$, $P = 70$W, $\xi = 1.5\sqrt{\frac{P}{MNN_\text{t}}}$, $\varepsilon = 1$).}\label{fig:SINR_SNR}
\end{figure}

\begin{figure}[htbp]
  \centering
  \includegraphics[width = 3.5 in]{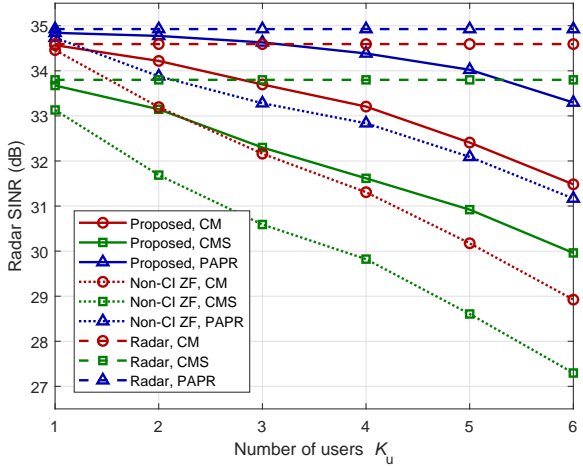}
\caption{\textcolor{black}{Radar output SINR versus the number of users $K_\text{u}$ ($\Gamma = 5$dB, $\theta_0 = 0$, $f_\text{d} = 0.3$, $P = 100$W, $\xi=1.5\sqrt{\frac{P}{MNN_\text{t}}}$, $\varepsilon=1$).}}\label{fig:SINR_Ku}
\end{figure}

\begin{figure}[!t]
\centering
\includegraphics[width = 3.5 in]{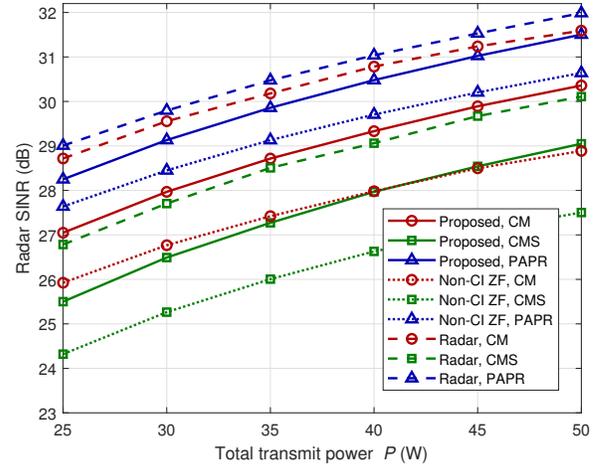}\vspace{0.1 cm}
\caption{Radar output SINR versus total transmit power ($\Gamma = 5$dB, $\theta_0 = 0$, $f_\text{d} = 0.3$, $\xi=1.5\sqrt{\frac{P}{MNN_\text{t}}}$, $\varepsilon=1$).}\label{fig:SINR_P}
\end{figure}

\begin{figure}[!t]
\centering
\includegraphics[width = 3.5 in]{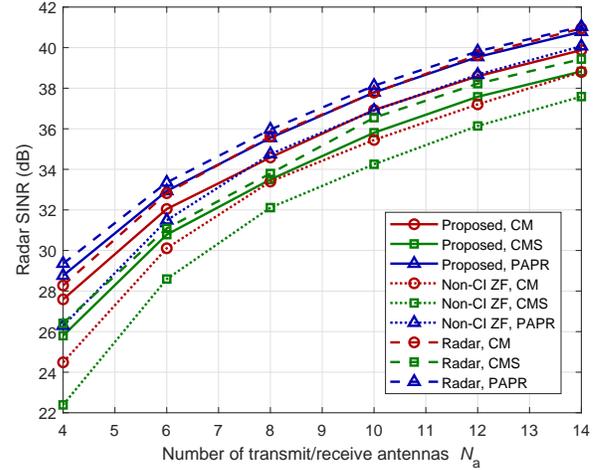}\vspace{0.1 cm}
\caption{\textcolor{black}{Radar output SINR versus the number of antennas $N_\text{a}$ ($\Gamma = 5$dB, $\theta_0 = 0$, $f_\text{d} = 0.3$, $P = 70$W, $\xi=1.5\sqrt{\frac{P}{MNN_\text{t}}}$, $\varepsilon=1$).}}\label{fig:SINR_Nt}
\end{figure}

\begin{figure}[!t]
\centering
\includegraphics[width = 3.5 in]{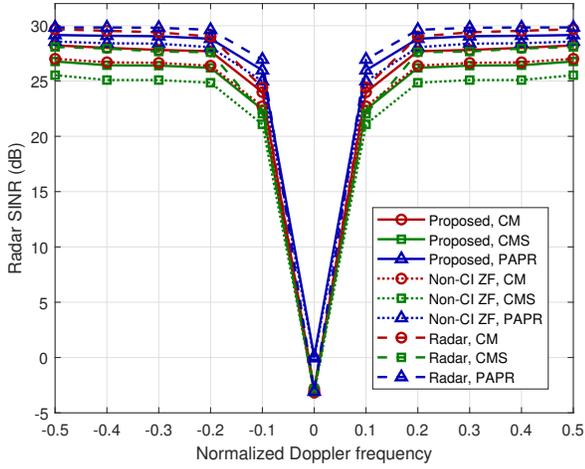}\vspace{0.1 cm}
\caption{Radar output SINR versus normalized Doppler frequency ($\Gamma = 5$dB, $\theta_0 = 0$, $P = 30$W, $\xi=1.5\sqrt{\frac{P}{MNN_\text{t}}}$, $\varepsilon=1$).}\label{fig:SINR_fd}
\end{figure}

\begin{figure}[!t]
\centering
\includegraphics[width = 3.5 in]{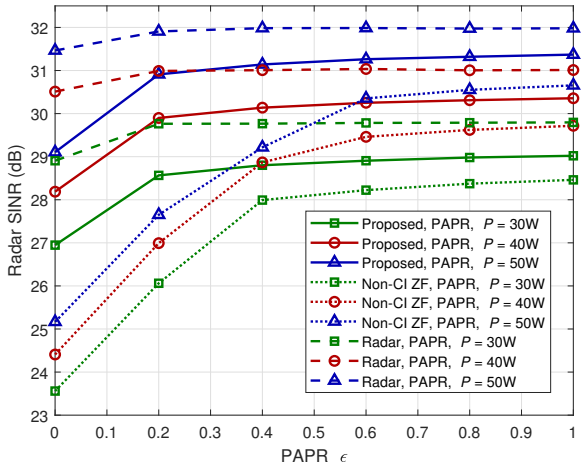}\vspace{0.1 cm}
\caption{Radar output SINR versus PAPR $\varepsilon$ ($\Gamma = 5$dB, $\theta_0 = 0$, $f_\text{d} = 0.3$).}\label{fig:PAPR}
\end{figure}

\begin{figure}[!t]
\centering
\includegraphics[width = 3.5 in]{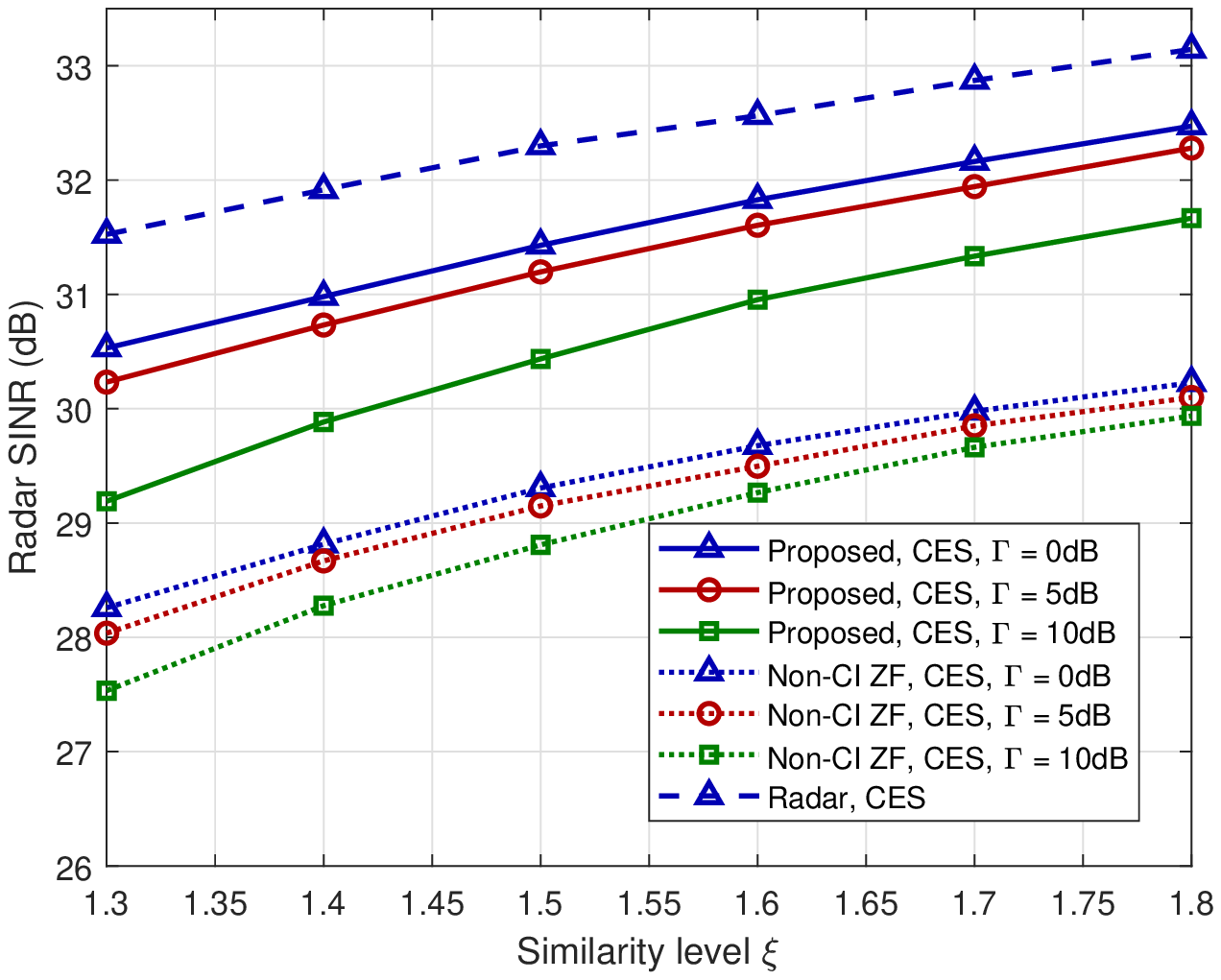}\vspace{0.1 cm}
\caption{Radar output SINR versus similarity level $\xi$ ($\theta_0 = 0$, $f_\text{d} = 0.3$, $P = 80$W).}\label{fig:similarity}
\end{figure}

The radar output SINR versus the communication QoS requirement $\Gamma$ is shown in Fig. \ref{fig:SINR_SNR}.
Not surprisingly, the radar output SINR achieved by the proposed MIMO-DFRC system decreases as the communication QoS requirement increases due to the trade-off between radar sensing performance and wireless communication QoS. As already noted, the PAPR-constrained waveform design achieves the highest radar SINR since it has the most relaxed waveform constraint, while the constant-modulus and similarity-constrained waveform design has the lowest radar SINR since it not only imposes the constant-modulus constraint on each transmit antenna, but also attempts to match the desired reference waveform in order to achieve other desired radar sensing properties in addition to the output SINR.
The performance of the constant-modulus waveform design lies in between.
In addition, we observe that the BS can provide 3 users with a communication QoS $\Gamma = 10$dB at the price of about 2dB in radar performance loss and the proposed CI-based approach has about 2dB performance improvement compared with the non-CI approach.
This phenomenon confirms the advantages of utilizing STAP and CI-based SLP techniques in MIMO-DFRC systems.
\textcolor{black}{In Fig. \ref{fig:SINR_Ku}, we present the radar output SINR versus the number of communication users.
The trade-off between the radar sensing performance and the wireless communication requirement can be clearly observed.
Moreover, the gap between the proposed CI-based schemes and the non-CI approaches becomes larger as $K_\text{u}$ increases, which demonstrates the advantage of the proposed CI-based approach in exploiting MUI in dense-user cases. }

Then, we illustrate the radar output SINR versus the total transmit power in Fig. \ref{fig:SINR_P}.
Clearly, a higher transmit power provides a larger radar output SINR.
Moreover, the performance relationship is the same as shown in Fig. \ref{fig:SINR_SNR} and similar conclusions can be drawn.
\textcolor{black}{We also present the radar output SINR versus the number of antennas $N_\text{a} = N_\text{t} = N_\text{r}$ in Fig. \ref{fig:SINR_Nt}. It is clear that adding antennas achieves better performance owing to the increased waveform diversity and higher beamforming gains.}
Fig. \ref{fig:SINR_fd} illustrates the radar output SINR for the proposed waveform designs under different normalized Doppler frequencies from $-0.5$ to $0.5$.
As expected, there is a significant SINR notch when the normalized Doppler frequency tends to zero, since the reflected signal from a slowly-moving target is difficult distinguish from the strong clutter returns.

In Fig. \ref{fig:PAPR}, we plot the radar output SINR versus the PAPR $\varepsilon$ for different levels of total transmit power for the proposed STAP-SLP-based DFRC approach and the standard MIMO radar scheme.
The radar output SINR increases with $\varepsilon$ since the power constraint on each transmit antenna becomes less strict.
Moreover, we observe that both the performance improvement and the gap between different schemes becomes smaller as $\varepsilon$ increases.
Fig. \ref{fig:similarity} illustrates the radar output SINR versus the similarity level $\xi$ with different communication QoS constraints.
The radar output SINR of all of the algorithms increases with increasing $\xi$ considering the trade-off between the radar output SINR metric and the other radar properties endowed by the reference waveform.

\begin{figure*}[!t]
\centering
\subfigure[Constant-modulus.]{
\begin{minipage}[t]{2.45 in}
\centering
\includegraphics[width = 2.45 in]{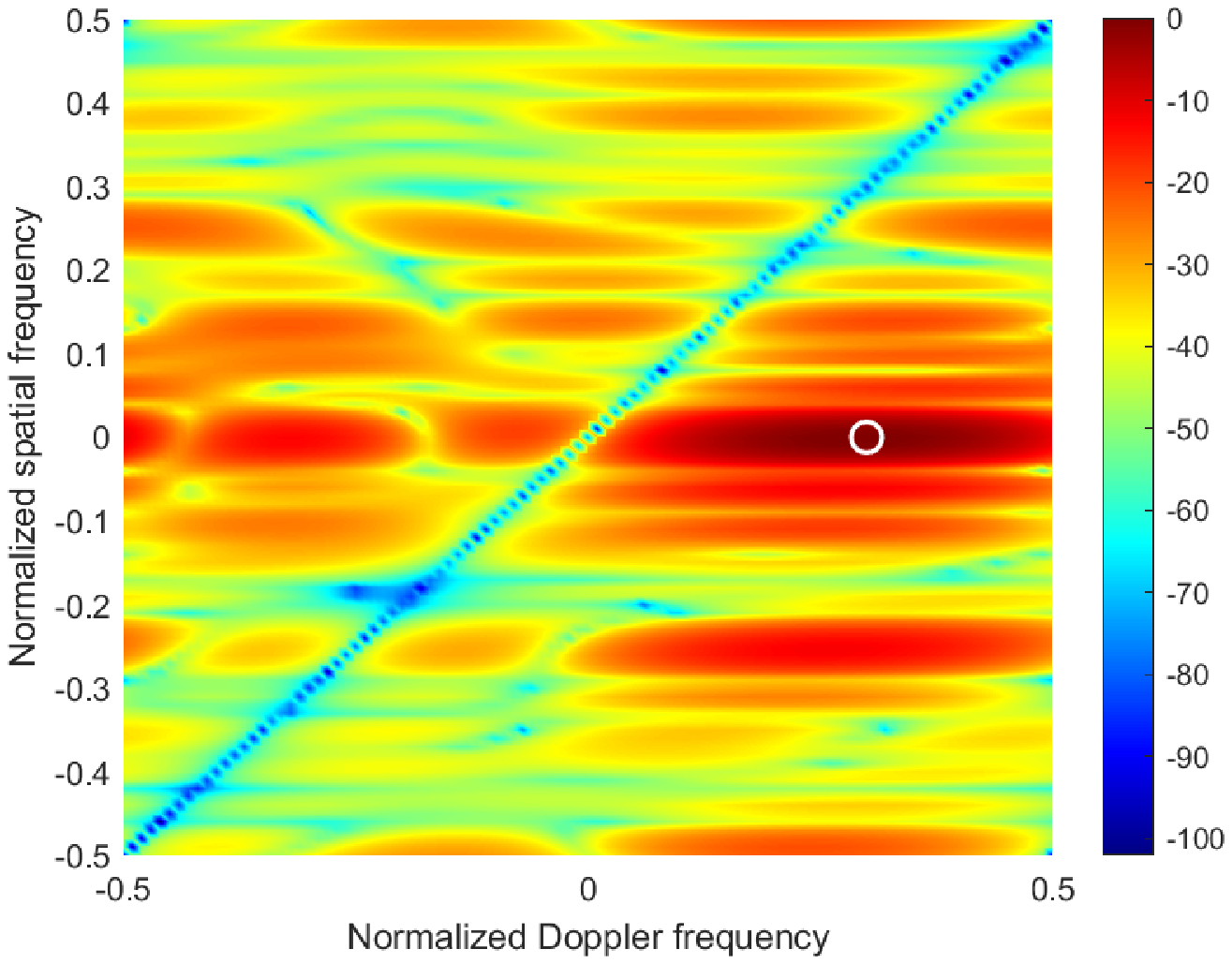}
\end{minipage}%
}%
\subfigure[PAPR.]{
\hspace{-0.2 cm}
\begin{minipage}[t]{2.45 in}
\centering
\includegraphics[width = 2.45 in]{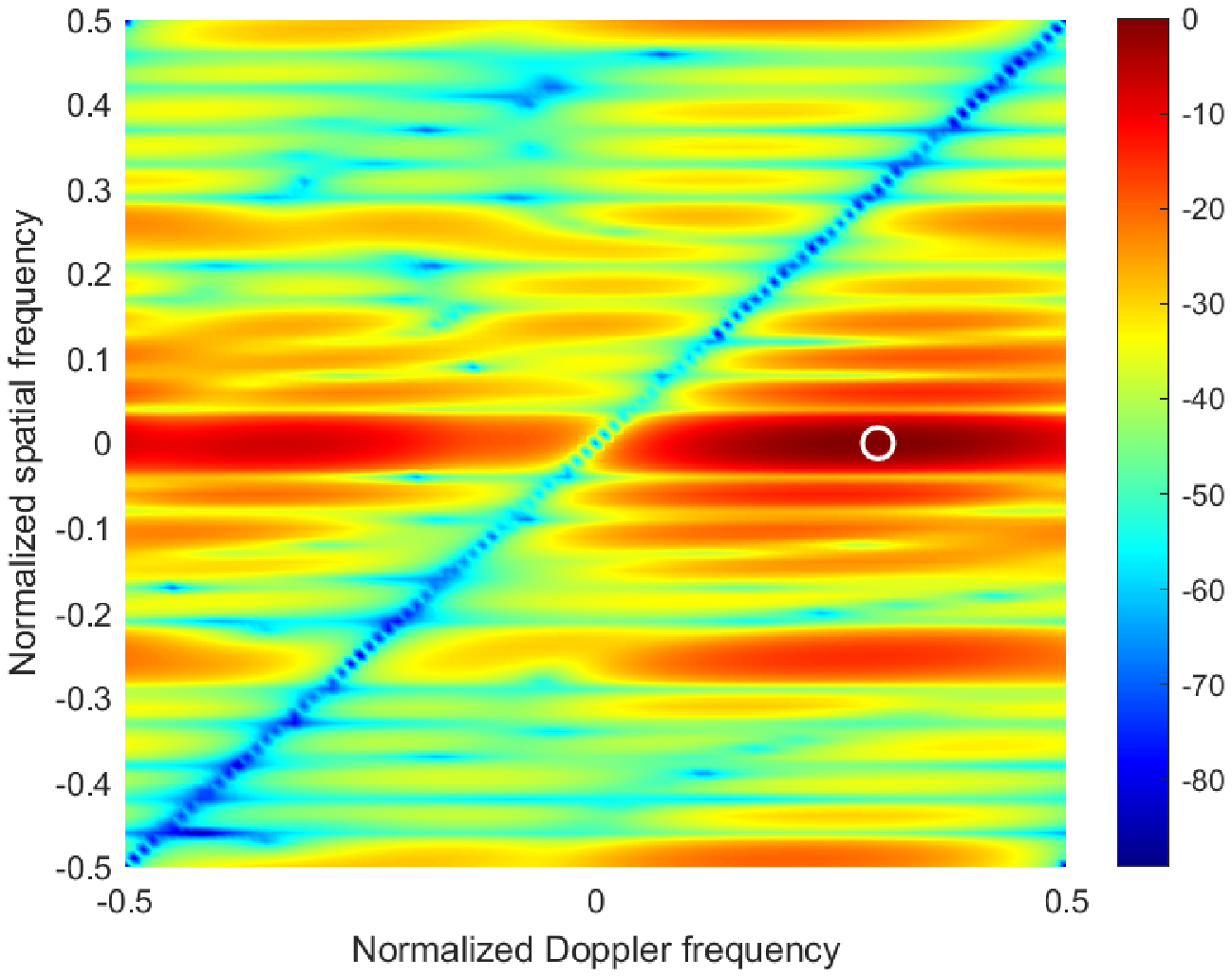}
\end{minipage}%
}%
\subfigure[Constant-modulus and similarity.]{
\hspace{-0.2 cm}
\begin{minipage}[t]{2.45 in}
\centering
\includegraphics[width = 2.45 in]{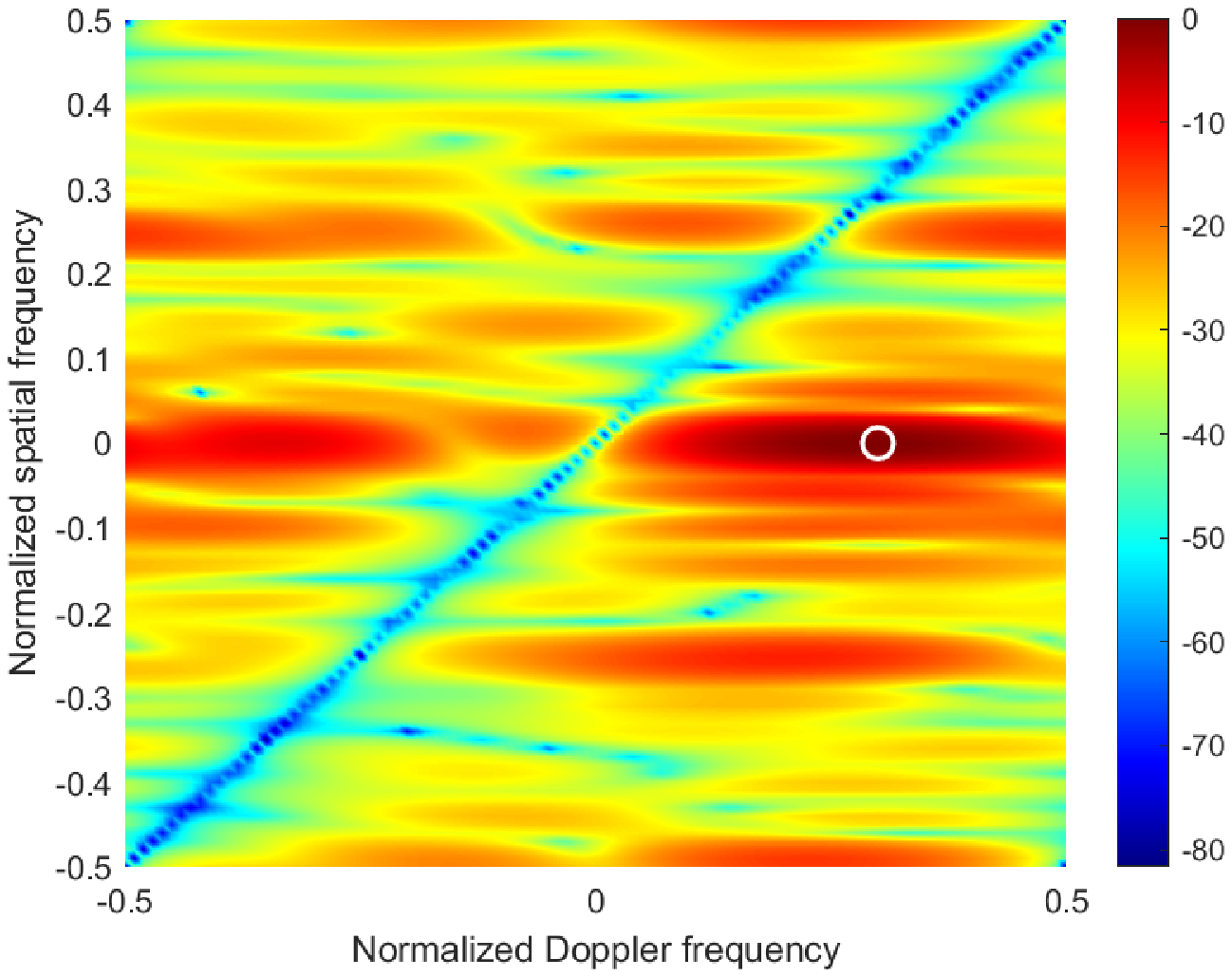}
\end{minipage}
}%
\centering
\caption{Space-time cross-ambiguity function under different waveform constraints ($\Gamma = 5$dB, $\theta_0 = 0$, $f_\text{d} = 0.3$, $P = 30$W, $\xi=1.5\sqrt{\frac{P}{MNN_\text{t}}}$, $\varepsilon=1$).}\label{fig:ambiguity_function}
\end{figure*}

Finally, in order to illustrate the capabilities of the designed DFRC waveform in target detection and clutter suppression, we plot the space-time cross-ambiguity functions of the three scenarios in Fig. \ref{fig:ambiguity_function}, where the white circles denote the target location (0 normalized spatial frequency and 0.3 normalized Doppler frequency) and where the number of clutter patches in each cell is set as $N_\text{c} = 100$.
The space-time cross-ambiguity function is defined by
\be
P_{\mathbf{w},\mathbf{x}}(f_\text{d},\theta) = \left|\mathbf{w}^H\overline{\mathbf{X}}\mathbf{u}(f_\text{d},\theta)\right|^2.
\ee
In these colormaps, it can be seen that the mainlobes are located at the target locations while the deep nulls span the clutter ridge.
This phenomenon verifies that the proposed STAP-SLP-based MIMO-DFRC system and waveform design can effectively suppress the signal-dependent clutter and achieve satisfactory target detection performance.

\section{Conclusions}\label{sec:conclusion}
\vspace{0.0 cm}

In this paper, we investigated STAP and SLP-based joint transmit waveform and receive filter designs for MIMO-DFRC systems.
The radar output SINR was maximized under different waveform and CI constraints that guarantee satisfactory communication QoS.
Efficient algorithms exploiting MM and neADMM methods were developed to solve the resulting complicated non-convex optimization problems.
Simulation examples demonstrated the advantages of utilizing STAP and CI-based SLP techniques to implement MIMO-DFRC, as well as the effectiveness of the proposed joint transmit waveform and receive filter design algorithms.
\textcolor{black}{Motivated by this initial work, we will further investigate other issues related to the implementation of STAP and SLP techniques in practical MIMO-DFRC systems, e.g., low-complexity algorithms, hardware imperfections, robust designs, etc. }

\end{document}